\documentclass[%
  reprint,
 twocolumn,
 notitlepage,
 amsmath,amssymb,
 aps,prb,
 ]{revtex4-1}

\usepackage{physics}
\usepackage{siunitx}
\usepackage{xspace}

\usepackage{graphicx}

\usepackage{dcolumn}
\usepackage{bm} 
\usepackage{bbm} 
\usepackage[colorlinks=true]{hyperref}

\usepackage{caption,setspace}
\usepackage{subcaption}
\usepackage{ragged2e}
\DeclareCaptionJustification{justified}{\justifying}
\newcommand{\RhoOp}{\ensuremath\hat{\rho}}
\newcommand{\CNOT}{\textsc{CNOT}\xspace}

\newcommand{\scinotation}[1]{\num[scientific-notation=true]{#1}}

\begin{document}

\title{Deterministic teleportation of a quantum gate between two logical qubits}
\author{K.S. Chou}\email{kevin.chou@yale.edu}
\author{J.Z. Blumoff}
\altaffiliation[Present address: ]{HRL Laboratories, LLC, 3011 Malibu Canyon Road, Malibu, California 90265, USA}
\author{C.S. Wang}
\author{P.C. Reinhold}
\author{C.J. Axline}
\author{Y.Y. Gao}
\author{L. Frunzio}
\author{M.H. Devoret}
\author{Liang Jiang}
\author{R.J. Schoelkopf}
\email{robert.schoelkopf@yale.edu}
\affiliation{%
	Department of Applied Physics and Physics Yale University, New Haven, Connecticut 06511, USA
}
\affiliation{%
	Yale Quantum Institute, Yale University, New Haven, Connecticut 06520, USA
}

\date{\today}



\maketitle

\textbf{
A quantum computer has the potential to efficiently solve problems that are intractable for classical computers.
Constructing a large-scale quantum processor, however, is challenging due to errors and noise inherent in real-world quantum systems.
One approach to this challenge is to utilize modularity---a pervasive strategy found throughout nature and engineering---to build complex systems robustly.
Such an approach manages complexity and uncertainty by assembling small, specialized components into a larger architecture.
These considerations motivate the development of a quantum modular architecture, where separate quantum systems are combined via communication channels into a quantum network \cite{Kimble2008, Monroe2014}.
In this architecture, an essential tool for universal quantum computation is the teleportation of an entangling quantum gate \cite{Gottesman1999,Eisert2000,Jiang2007}, a technique originally proposed in 1999 which, until now, has not been realized deterministically.
Here, we experimentally demonstrate a teleported controlled-NOT (\CNOT) operation made deterministic by utilizing real-time adaptive control. 
Additionally, we take a crucial step towards implementing robust, error-correctable modules by enacting the gate between logical qubits, encoding quantum information redundantly in the states of superconducting cavities \cite{Ofek2016}.
Such teleported operations have significant implications for fault-tolerant quantum computation \cite{Gottesman1999}, and when realized within a network can have broad applications in quantum communication, metrology, and simulations \cite{Duan2001, Kimble2008, Monroe2014}.
Our results illustrate a compelling approach for implementing multi-qubit operations on logical qubits within an error-protected quantum modular architecture. 
}

\begin{figure*}[tp]
	\centering{
		\phantomsubcaption\label{subfig:mod_arch}
		\phantomsubcaption\label{subfig:module}
		\phantomsubcaption\label{subfig:teleported_gate}
		\phantomsubcaption\label{subfig:device}
	}
	\includegraphics{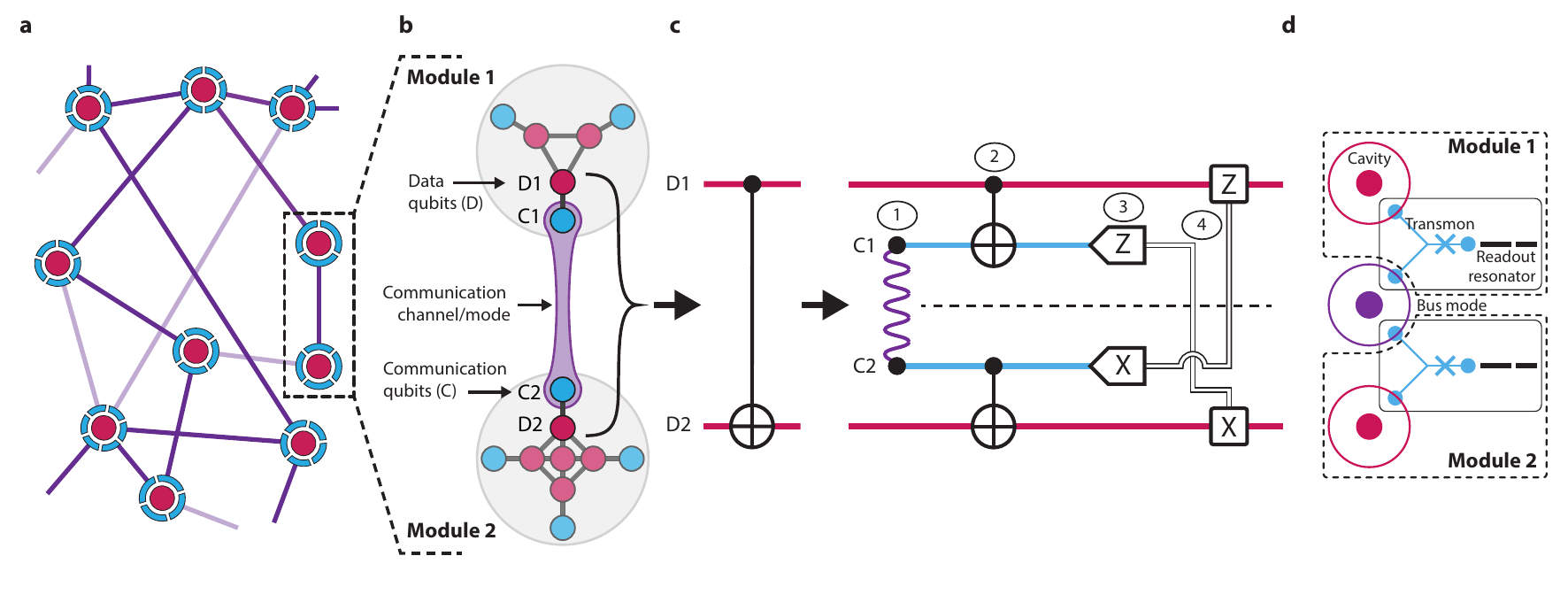}
	\caption{\label{fig:modules_cnot_device}{
			\textbf{Construction of a modular architecture and teleported \CNOT gate}.
			\subref{subfig:mod_arch}, Network overview of the modular quantum architecture.
			Modules are represented as nodes of a quantum network and are composed of: data qubit(s) (magenta) and communication qubit(s) (cyan). Coupling between modules is generated through potentially reconfigurable communication channels that may be enabled (dark purple line) or disabled (light purple line).
			\subref{subfig:module}, Quantum modules.
			Each module houses a small quantum processor capable of high fidelity operations among data qubits and communication qubits.
			In our experiment, we create two modules each consisting of one data qubit (D1 and D2) and one communication qubit (C1 and C2). 
			\subref{subfig:teleported_gate}, Teleported \CNOT circuit between D1 and D2.
			The teleported \textsc{CNOT} circuit requires (1) entanglement between C1 and C2 (purple meander), (2) local operations, (3) measurement of C1 and C2, and (4) classical communication (double lines) and feedforward operations.
			\subref{subfig:device}, Experimental realization (schematic top view) in a 3D cQED implementation. Each module consists of a data qubit defined as a coaxial $\lambda / 4$ 3D cavity (magenta), a communication qubit defined as a Y-shaped transmon qubit (cyan), and a Purcell-filtered, quasi-planar, $\lambda / 2$ stripline readout resonator (black).
			In this experiment, the two modules are linked by an additional mode realized as a coaxial $\lambda / 4$ 3D cavity (purple) that serves as a bus mode. 
			Additional details are provided in the Supplementary Information.
	}}
\end{figure*}

The quantum modular architecture is a distributed network (\autoref{subfig:mod_arch}) of modules that communicate with one another through quantum and classical channels. 
Each module is a small quantum processor that is composed of two separately optimized subsystems (\autoref{subfig:module}): 
first, data qubits that function as quantum memories and are logically encoded to be error correctable; 
and second, communication qubits that mediate interactions between different modules through distributed entanglement.
This architecture uses distributed entanglement as a vital quantum resource for performing multi-qubit operations between data qubits. 
These operations are enabled through teleportation and allow the data qubits to be well-isolated, offering a systematic path for minimizing crosstalk and residual interactions across the entire network even while scaling the system.
So far, elementary quantum networks have demonstrated the transmission of quantum information and the generation of entanglement between communication qubits \cite{Ritter2012,Bernien2013,Hucul2014,Narla2016}.
It will be necessary to implement entangling operations between logical data qubits to perform universal quantum computation using these networks. 

In contrast to conventional approaches that use direct interactions, the modular quantum architecture will require quantum teleportation to enact entangling operations \cite{Bennett1993,Gottesman1999}.
Teleportation has been used in a variety of platforms to transfer a quantum state between two remote systems \cite{Bouwmeester1997,Furusawa1998,Riebe2004,Barrett2004,Sherson2006,Olmschenk2009,Ma2012,Steffen2013}.
Expanding on this technique, the teleportation of a two-qubit quantum gate implements a unitary operation between two unknown states with a protocol that circumvents the necessity for direct interaction between the two data qubits \cite{Gottesman1999,Jiang2007} (\autoref{subfig:teleported_gate}). 
Instead, these teleportation-based protocols utilize a previously prepared entangled state of the communication qubits, local operations within each module, and classical communication between modules \cite{Gottesman1999,Eisert2000,Jiang2007}.
Previously, similar protocols have been demonstrated between two physical data qubits without real-time classical communication \cite{Huang2004,Gao2010,K.2017}, where the desired operation is extracted probabilistically through postselection.
However, to avoid excessive overhead and to make the modular approach scalable, it is crucial to perform these teleported gates deterministically.



In our work, we demonstrate a teleported \CNOT gate that is both deterministic and operates on logically-encoded data qubits. 
We implement two modules that each consist of a superconducting microwave cavity as the data qubit and a transmon as the communication qubit.
Here, we generate entanglement between communication qubits via a local quantum bus that individually couples to each communication qubit.
Our implementation can be adapted in the future to incorporate schemes for generating remote entanglement \cite{Roch2014,Narla2016}, necessary for a scalable quantum modular architecture.
Here, we use a hardware-efficient approach \cite{Ofek2016,Michael2016} to logically encode each data qubit within the states of a long-lived cavity mode.
Importantly, despite the added complexity of our logical encoding, we implement high-fidelity control over both data and communication qubit within each module.
Using the teleported \CNOT gate combined with real-time adaptive control, we generate a Bell state and characterize the logical quantum process, thus validating our entangling operation on logical qubits.
Our physical implementation capitalizes on highly coherent and controllable elements from the three-dimensional (3D) circuit quantum electrodynamics (cQED) platform.
Each module (\autoref{subfig:device}) consists of a high-Q 3D electromagnetic cavity \cite{Reagor2016} as the data qubit, a transmon qubit as the communication qubit, and a Purcell-filtered, low-Q stripline resonator \cite{Axline2016} as the readout for the transmon qubit.
The transmon qubit is capacitively coupled to both the data qubit and readout resonator. 
Notably, we achieve data qubit coherences (${\sim}\SI{1}{ms}$) that are around three orders of magnitude greater than the measurement time (${\lesssim}\SI{1}{\micro\second}$), enabling both quantum information storage as well as fast measurement within a single package (\autoref{subfig:device}, also see Supplementary Information).
In this experiment, the communication channel is implemented as an additional cavity mode and functions as a quantum bus (hereafter, ``bus''), coupling individually to both communication qubits.
Importantly, though we utilize this local mode to link the two modules, the two data qubits have an immeasurably-small direct coupling, bounded to be at least an order of magnitude smaller than the smallest decay rate in our system (Supplementary Information).
Therefore, despite the physical proximity between the two modules, our two data qubits are effectively non-interacting, demonstrating the same isolation distinctive of remote modular architectures.
Additionally, our entire device exhibits low readout crosstalk, which is critical for the teleported gate and is a characteristic property of independent modules (Supplementary Information).

\begin{figure}[tp]
	\centering{\phantomsubcaption\label{subfig:logical_bloch}
		\phantomsubcaption\label{subfig:truth_table}
	}
	\includegraphics{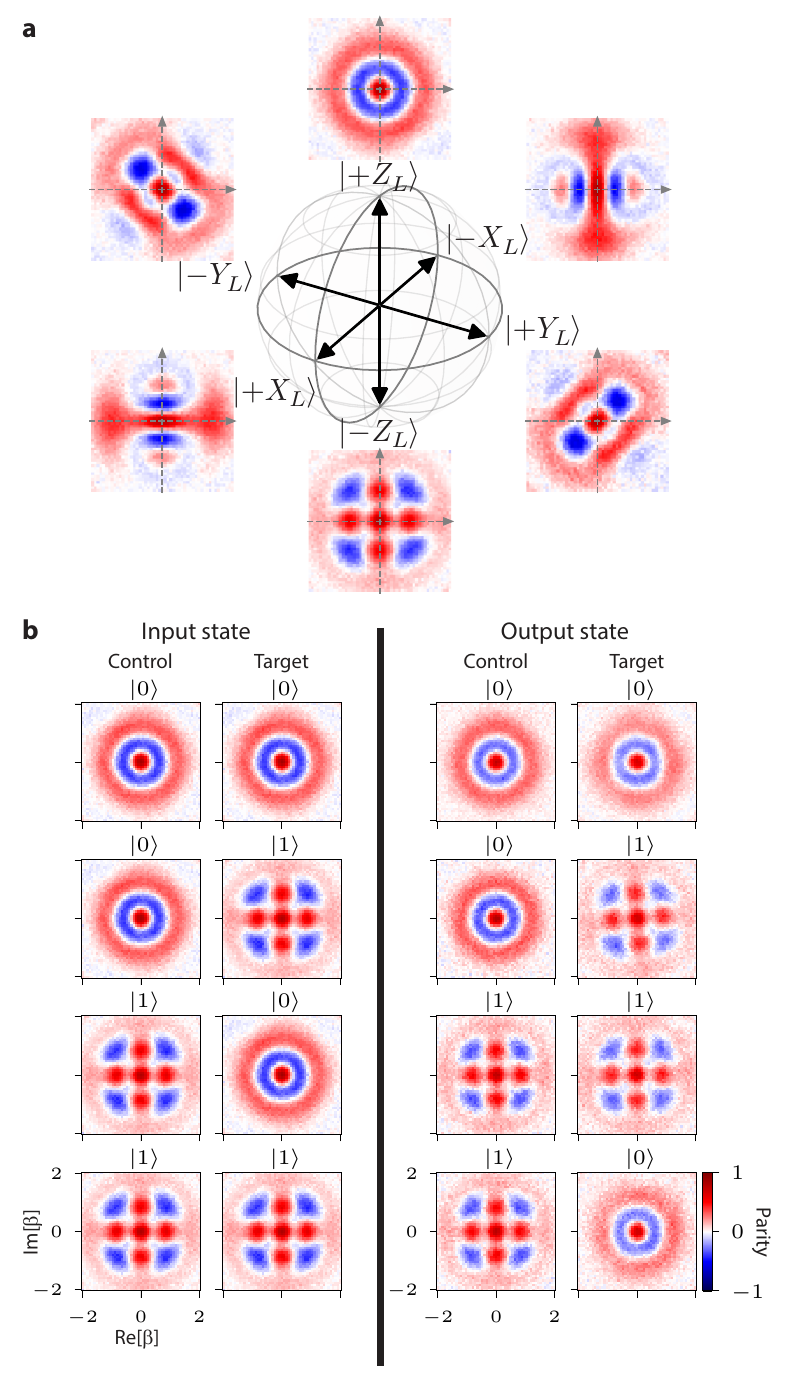}
	\caption{\label{fig:logical_qubit_truth_table}{
			\textbf{Logical data qubit encoding and \textsc{CNOT} truth table.}
			\subref{subfig:logical_bloch}, Logical Bloch sphere for the binomial code encoding.
			The data qubit is logically encoded in the binomial code basis and the Wigner function for each of the six cardinal states $\lbrace {\pm}\hat{Z}_L, {\pm}\hat{X}_L, {\pm}\hat{Y}_L \rbrace$ is shown.
			\subref{subfig:truth_table}, Teleported \CNOT truth table. 
			The left two columns show experimental Wigner functions illustrating all four logical computational states as input states, and the right two columns show the extracted Wigner functions after performing the teleported \textsc{CNOT} operation, illustrating the correct classical behavior of the gate. 
	}}
\end{figure}

The high-dimensional cavity modes that define our data qubits allow for a wide range of encodings.
For our demonstration of the teleported \CNOT, we have chosen to logically-encode each of the data qubits as one of the recently-developed bosonic binomial quantum codes \cite{Michael2016},
\begin{equation}
\ket{0_L} = \ket{2},\qquad \ket{1_L} = \frac{1}{\sqrt{2}}\left(\ket{0} + \ket{4}\right),
\end{equation}
specified in the photon-number basis of the cavity.
In the Supplementary Information, we also present results using the $\ket{0}$ and $\ket{1}$ Fock basis to highlight that our protocol is flexible to different data qubit encodings.
For the binomial code, the two basis states have even photon-number parity. When this logical qubit functions as a quantum memory, the dominant error channel for this system is single-photon loss, an error channel that transforms the basis states into states of odd photon-number parity. Importantly, in this encoding the quantum information is still preserved in this error subspace, such that photon-number parity measurements---which can implemented with high fidelity in our system \cite{Ofek2016}---can be used in an error-correction protocol to detect and correct single-photon loss events in the cavity.
To illustrate the logical Bloch sphere, we prepared the six cardinal states and characterized each state by measuring the Wigner function of the data qubit (\autoref{subfig:logical_bloch}).
The Wigner function provides not only a strikingly visual representation of the logical qubit state, but also completely specifies the underlying cavity state, a capability analogous to full state tomography of the constituent physical qubits that compose a logical qubit.

The teleported \CNOT starts with the generation of entanglement in the communication qubits to create a communication channel between the two modules (Step 1 in \autoref{subfig:teleported_gate}).
In our implementation, we use the Bell state $\ket{\Psi^+} = \left( \ket{ge} + \ket{eg} \right) /\sqrt{2}$, though any chosen maximally-entangled state is acceptable, requiring only small modifications to later steps of the protocol. 
The state is generated by performing a resonator-induced phase (RIP) gate \cite{Paik2016} on the bus and single-qubit rotations on the communication qubits.
The Bell state generation occurs while the data qubits store quantum information; the static dispersive interaction, if not accounted for, will naturally entangle the data and communication qubits.
Because it is necessary for the two qubits within each module to be disentangled at the end of this step, we modify our Bell pair generation protocol and implement a refocused RIP sequence \cite{Paik2016} to echo away this unwanted interaction independent of the data qubit encoding scheme.
An important consequence of the Bell state generation protocol is that the dispersive interaction induces a known, deterministic reference frame shift on each of the data qubits; these are accounted for in subsequent steps of the teleported gate protocol (Supplementary Information).
Using this modified sequence, we have generated a Bell pair between the communication qubits in ${\sim}\SI{680}{ns}$ with state fidelity of $\left(97\pm 1\right)\%$ as determined from quantum state tomography (Supplementary Information).

Next, local operations performed within each module entangle the data and communication qubits (Step 2 in \autoref{subfig:teleported_gate}).
Our local operations are implemented using optimal control techniques which enable universal quantum control between the data and communication qubits \cite{Heeres2017}.
We generate all of our local operations with pulse lengths between $\SI{1}{\us}$ and $\SI{2}{\us}$.
Characterization of these logical operations yields single data qubit and two-qubit (between the data and communication qubits) gate fidelities of ${\sim}97\%$ and ${\sim}94\%$, respectively (Supplementary Information).

After the entangling local operations, we perform measurements on the communication qubits (Step 3 in \autoref{subfig:teleported_gate}), thereby effecting a unitary operation between only the two data qubits.
It is critical that the measurements do not reveal information about the state of the data qubits. 
In our protocol, this is accomplished by individual measurements of the communication qubits in the $\hat{Z}$ and $\hat{X}$ bases, which lead to four uniformly distributed outcomes. 
Each outcome heralds a unitary operation between the two data qubits that is a \CNOT gate up to single-qubit operations. 
As a result, high-fidelity measurements are necessary to correctly determine the particular operation enacted on the data qubits.
In our system, we achieve single-shot state assignment fidelities of the communication qubits around $99\%$ (Supplementary Information).

Finally, ensuring that the protocol implements the desired \CNOT operation independent of measurement outcome requires classical communication and feedforward operations (Step 4 in \autoref{subfig:teleported_gate}).
Two classical bits of information are needed to communicate measurement results between modules.
This information is used to apply feedforward operations, transforming the protocol into a deterministic operation and thus completing the teleportation.
In our experiment, it is required that the measurements be non-destructive to the communication qubits as they are used for subsequent steps of our protocol.
For our protocol, these measurements also induce a conditional reference phase shift on the data qubits dependent on measurement outcome (Supplementary Information). 
Tracking these phases accurately is critical for all subsequent operations on the data qubits. 
To enable both the measurements and the feedforward, we employ a real-time controller \cite{Ofek2016} to orchestrate quantum programs for our experiment, combining control, measurement, state estimation, and feedfoward in a single integrated system. 
For every experimental run, this controller handles the distribution of classical information between the two modules, dynamically updating the reference phases and applying the appropriate feedforward operations, all within a fraction of the lifetime (${\sim}1\%$) of the communication qubits.
We have independently analyzed the measurement and feedforward process to have a combined fidelity of ${\sim}97\%$, excluding the data qubit operations (Supplementary Information).


Therefore, by consuming a shared entangled pair and communicating two classical bits of information, this procedure effects a \CNOT operation between the data qubits without requiring a unitary operation between the two modules after the generation of the shared entangled pair.
Having demonstrated all of the elements necessary for realizing the teleported \CNOT gate, we characterized the full two-qubit gate through a series of four separate analyses.

In the first analysis, we verified the classical behavior of the gate by generating a truth table for the set of computational states.
We prepared the data qubits each of the four states $\lbrace\ket{0_L 0_L}$, $\ket{0_L 1_L}$, $\ket{1_L 0_L}$, $\ket{1_L 1_L} \rbrace$ and enacted the teleported \textsc{CNOT} on each, ideally leading to the output states: $\lbrace\ket{0_L 0_L}$, $\ket{0_L 1_L}$, $\ket{1_L 1_L}$, $\ket{0_L 1_L} \rbrace$.
We extracted both the input and output states by measuring Wigner functions for each data qubit.
Our results (\autoref{subfig:truth_table}) provide qualitative validation of the teleported \CNOT on the computational basis states.

In the second analysis, we have demonstrated that it is a distinctly quantum operation by using the teleported \CNOT to generate entanglement between two logical qubits.
We prepared the data qubits in the separable initial state $\ket{\psi_\text{in}} = \left( \ket{0_L} + \ket{1_L} \right) \ket{0_L} / \sqrt{2}$ and performed the gate. The ideal output state is the Bell state $\ket{\Phi^+_L} = \left( \ket{0_L 0_L} + \ket{1_L 1_L} \right) / \sqrt{2}$.
We verified that our teleported \CNOT generates this logical qubit Bell pair using two separate methods, which together highlight our ability to characterize the data qubits both on a logical level (i.e. the encoded two-dimensional subspace) as well as on a physical level (i.e. the multi-dimensional cavity state).

\begin{figure}[tp]
	\centering{
		\phantomsubcaption\label{subfig:hybrid_wigner}
		\phantomsubcaption\label{subfig:hybrid_bell_qst}
	}
	\includegraphics{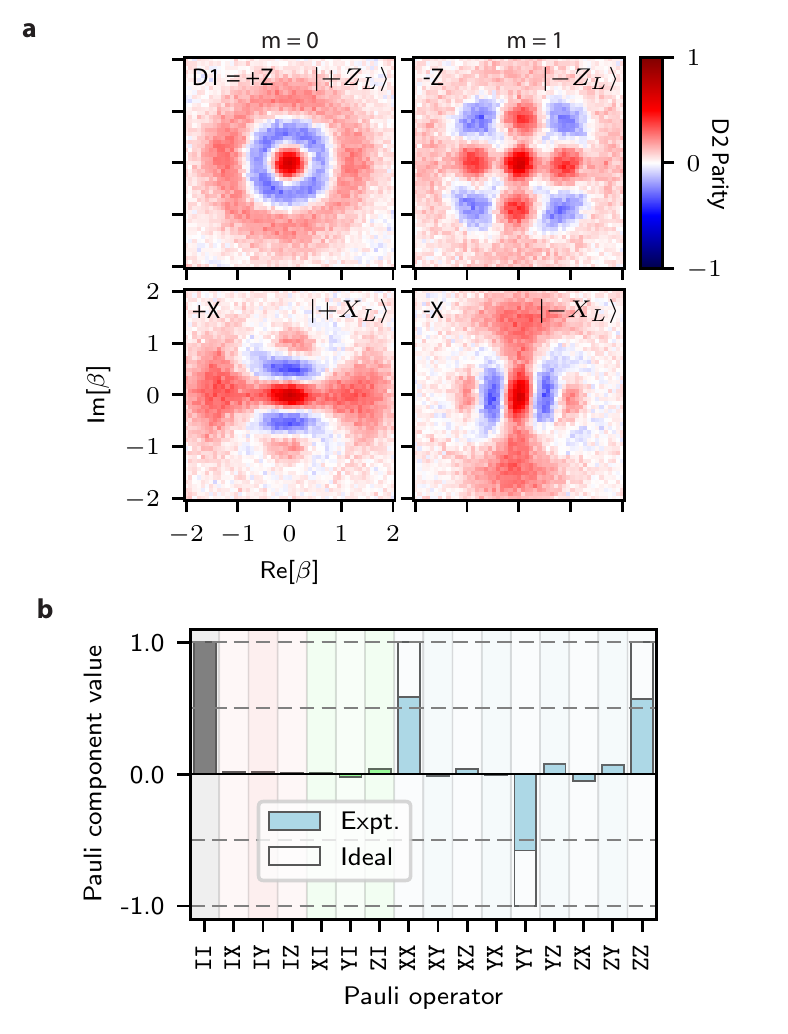}
	\caption{ \label{fig:hybrid_wigner}{
			\textbf{Generation of a logical Bell state.} 
			\subref{subfig:hybrid_wigner}, Quantum correlations of a logical Bell state.
			The logical Bell state $\ket{\Phi^+_L}$ is first created using the teleported \CNOT gate. 
			The control qubit is measured in either the logical $\hat{Z}_L$ basis (top) and $\hat{X}_L$ basis (bottom), and Wigner tomography is performed on the the target qubit conditioned on the measurement result, $m = 0$ (left) and $m = 1$ (right). 
			Correlations between the measurement result and measured state signal the generation of an entangled state between D1 and D2.
			\subref{subfig:hybrid_bell_qst}, Logical state tomography. 
			After generating $\ket{\Phi^+_L}$, logical qubit tomography is performed on both the control and target qubit. The reconstructed state, represented in the Pauli basis, confirms the teleported \CNOT has generated the target Bell state.
	}}
\end{figure}

In the first method, we performed a pair of experiments to show that the state exhibits quantum correlations.
Given the target state $\ket{\Phi_L^+}$, when we measure the control qubit in the logical $\hat{Z_L}$ basis and find it in $\ket{0_L}$ ($\ket{1_L}$), we expect the target qubit to be $\ket{0_L}$ ($\ket{1_L}$).
We enacted the logical $\hat{Z_L}$ measurement and, conditioned on the result, performed physical-qubit tomography on the target data qubit by measuring its Wigner function.
Experimentally, the logical measurement of the control data qubit is accomplished by first, decoding the state of the data qubit onto the communication qubit and then, measuring the desired observable on the communication qubit (Supplementary Information).
As expected, we observed strong $\hat{Z}$-correlations between the control and target data qubits (\autoref{subfig:hybrid_wigner}, top).
Next, we rotated the measurement basis and performed $\hat{X_L}$ measurements of the control data qubit. 
Conditioned on the control data qubit in $\ket{{\pm}X_L} = \left( \ket{0_L} \pm \ket{1_L} \right)/ \sqrt{2}$, we experimentally found the target data qubit to be in the expected state $\ket{{\pm}X_L}$ (\autoref{subfig:hybrid_wigner}, bottom), thus establishing $\hat{X}$-correlations between the two data qubits.
These two complementary experiments confirm the non-classical nature of the experimental logical Bell state and indicate that our gate produced a non-separable two-qubit state.

In the second method, we analyzed the joint state within the logical subspace of the two data qubits by performing quantum state tomography.
Quantum state tomography is performed using the same decoding technique as the logical qubit measurement discussed above.
We reconstructed the two-qubit state in the Pauli basis (\autoref{subfig:hybrid_bell_qst}), extracting a state fidelity $\mathcal{F}_\text{Bell} = \left( 68 \pm 1 \right)\%$ and concurrence $\mathcal{C} = \left( 0.37 \pm 0.01\right)$, exceeding the threshold for a classically correlated state.
These quantities include imperfections associated with logical state preparation and decoding operations, which together contribute about $6\%$ infidelity for each data qubit.
Importantly, using the teleported \textsc{CNOT}, we have generated a Bell state between logical qubits encoded as multi-photon states that, from inspection of the reconstructed density operator, has dominant two-qubit Pauli components (e.g. two-qubit parity, $\expval{ZZ} = 0.57$) and near-zero single-qubit Pauli components (e.g. single-qubit parities, $\expval{IZ} = 0.01$ and $\expval{ZI} = 0.04$).

\begin{figure*}[tp]
	\centering{
		\phantomsubcaption\label{subfig:qst_feedforward}
		\phantomsubcaption\label{subfig:qpt}
	}
	\includegraphics{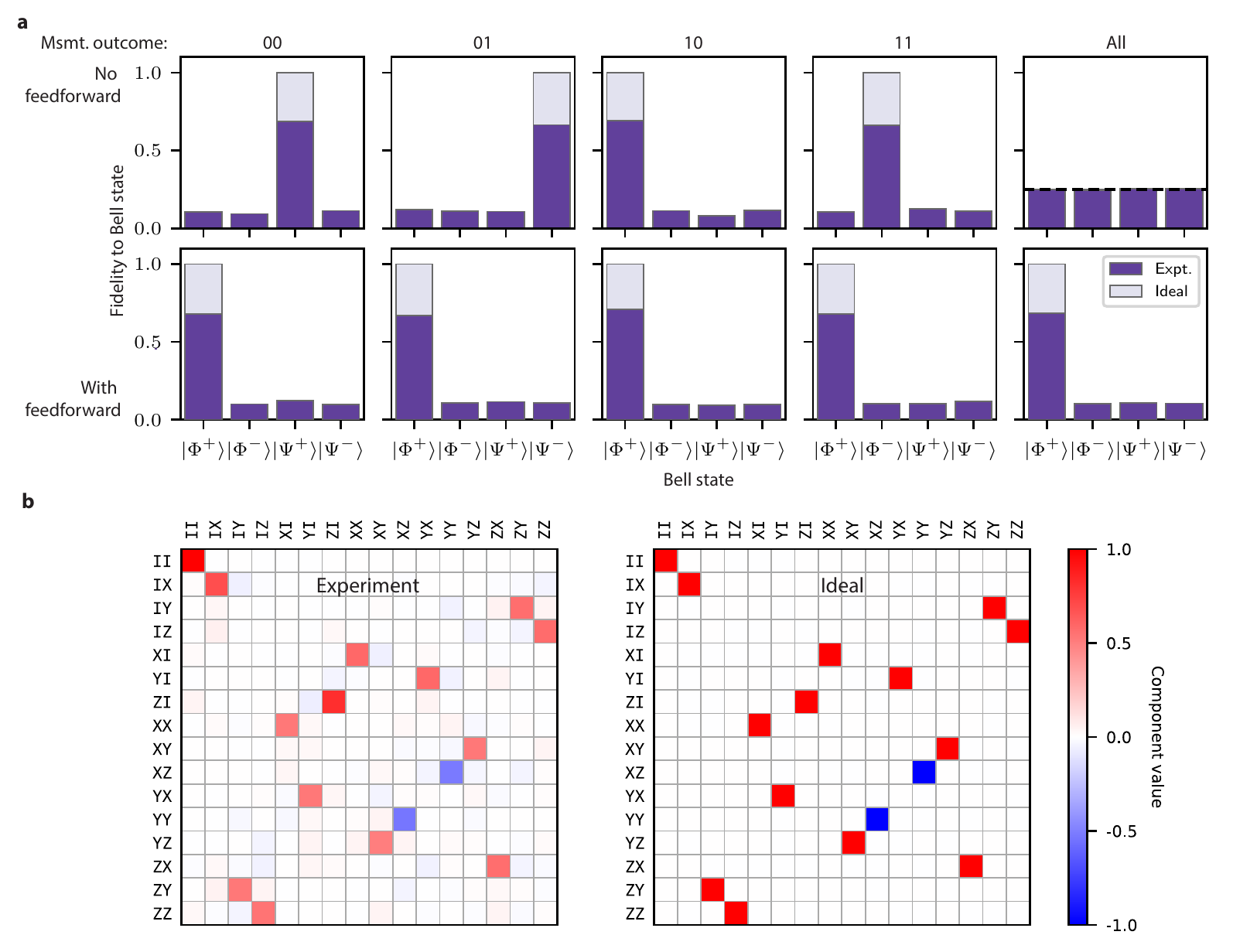}
	\caption{\label{fig:qpt}{
			\textbf{Demonstration of a deterministic teleported \textsc{CNOT} gate.}
			\subref{subfig:qst_feedforward}, Effect of feedforward operations.
			The teleported \textsc{CNOT} is applied to $\ket{\psi_\text{in}} = \left( \ket{0_L} + \ket{1_L} \right) \ket{0_L} / \sqrt{2}$ and the fidelity of the resulting state to each of the four Bell states is extracted. 
			Feedforward operations are not applied (top), and each measurement outcome $\lbrace00$, $01$, $10$, $11\rbrace$ results in a different Bell state. If all measurement results are compiled together, the resulting state is completely mixed (All).
			On the other hand, if the feedforward operations are applied (bottom), then the correct state $\ket{\Phi^+}$ is found for every measurement outcome. 
			\subref{subfig:qpt}, Quantum process tomography of the teleported \textsc{CNOT} gate.
			We represent the quantum process $\mathcal{R}_\CNOT$ in the Pauli transfer representation, in which the process map is expressed in the Pauli basis: $\vec{P}_{out} = \mathcal{R}_{\textsc{CNOT}} \vec{P}_{in}$, given input and output state Pauli vectors $\vec{P}_{in,out}$ (Supplementary Information).
			Agreement between the experimentally reconstructed (left) and ideal (right) process indicates the successful implementation of a deterministic teleported \CNOT gate.
	}}
\end{figure*}

Our implementation of the teleported gate as a deterministic operation requires reliable classical communication and feedforward operations.
In this third analysis, we investigated the importance of these elements by performing the previously-described data qubit entanglement sequence without applying the feedforward operations (Step 4 in \autoref{subfig:teleported_gate}).
Instead, we recorded the measurement outcomes and extracted four conditioned output states.
Without these feedforward operations, each measurement outcome $\lbrace 00, 01, 10, 11 \rbrace$ ideally occurs with probability $1/4$ and heralds one of four Bell states: $\lbrace \ket{\Psi^+_L}, \ket{\Psi^-_L}, \ket{\Phi^+_L}, \ket{\Phi^-_L}\rbrace$, where $\ket{\Psi^\pm} = \left( \ket{01} \pm \ket{10} \right) / \sqrt{2}$ and $\ket{\Phi^\pm} = \left( \ket{00} \pm \ket{11} \right) / \sqrt{2}$.
Our results (\autoref{subfig:qst_feedforward}; top, first four panels) are consistent with the ideal, save for reduced contrast, and we extracted conditioned fidelities of $\lbrace 69\%, 66\%, 69\%, 66\%\rbrace$ and outcome frequencies of $\lbrace 0.25, 0.26, 0.24, 0.25\rbrace$.
Crucially, the fact that we generated different Bell pairs indicates that each conditional operation is a \CNOT gate up to single qubit operations.
Without real-time knowledge of these measurement outcomes, these states will all add incoherently, resulting in a completely mixed state where all information has been lost, (\autoref{subfig:qst_feedforward}; top, All).
If we instead postselected on the measurement outcomes, the operation is left as a probabilistic two-qubit gate, achieving the target operation only $1/4$ of the time (\autoref{subfig:qst_feedforward}; top, measurement outcome 10).
Therefore, it is only when we combine real-time classical communication and feedforward that we can implement a deterministic teleported operation that performs the correct process independent of measurement outcome (\autoref{subfig:qst_feedforward}, bottom).

Finally, for the fourth analysis we fully characterized the logical process for the teleported \CNOT gate.
We performed quantum process tomography on the two logical qubits and our reconstructed process matrices show qualitative agreement with the expected process (\autoref{subfig:qpt}).
From the experimental reconstruction, we calculate a process fidelity of $\mathcal{F}_\text{pro} = \left( 68 \pm 2\right)\%$ without accounting for logical encoding or decoding steps that subtract from the extracted gate fidelity. 
With these corrections included (Supplementary Information), we infer a process fidelity of $\mathcal{F}_\text{gate} = \left(79 \pm 2 \right)\%$ for our teleported \CNOT gate.
To evaluate the experimental performance of the teleported gate, we assembled an error budget that combines the infidelity of each element of the gate, accounting for the known imperfections of our system.
From this analysis (Supplementary Information), we expect a gate fidelity of $\mathcal{F}_\text{thy} \approx \left(84 \pm 3\right)\%$, which is consistent with experimental results.
This indicates that other nonidealities, such as residual interactions or imperfect system characterization, are smaller effects in our system.
Indeed, as logical qubit operations---such as our teleported gate---are typically constructed from several distinct elements, it is necessary to experimentally verify that the compiled operation does not introduce unexpected errors. These considerations motivate efforts to construct and validate logical quantum systems both to reveal experimental nonidealities and to advance computational capabilities.

Building on our results, there exist well-defined prescriptions to further improve each element of the teleported gate.
In the present implementation, the dominant source of infidelity is the finite coherence of both communication qubits ($T_2 \approx \SI{15}{\micro\s}$), which accounts for about $70\%$ of the total gate error (Supplementary Information).
Though increasing transmon coherence is a straightforward approach to improve gate performance, a more robust solution may be to pursue alternate implementations of the communication qubit itself. 
For example, replacing the transmon qubit with a high-Q cavity will directly address this dominant source of infidelity in the present experiment. 
Furthermore, recent work demonstrating local operations between two cavities \cite{Rosenblum2017} illustrates that a module containing error-correctable data and communication qubits can be realized without sacrificing quantum control.
In addition to local operations, the performance of the teleported gate also depends on the quality of the shared entangled pair and communication qubit measurements.
Since the shared entangled pair can be prepared prior to the teleported operation, the gate is agnostic to how the entanglement is generated. Therefore, the protocol can take advantage of a variety of approaches, including deterministic \cite{Cirac1997} and probabilistic \cite{Barrett2005} schemes, and should benefit from entanglement purification protocols \cite{Bennett1996,Jiang2007}.
Measurements directly impact the performance of teleportation-based protocols, and strategies that boost measurement fidelity such as robust, repeated measurements \cite{Jiang2007} may be readily integrated into the teleported \CNOT gate.
Finally, since our implementation is compatible with various logical data qubit encodings, it may be possible to tailor the encoding to account for the dominant error channels associated with the gate (in the current implementation, codespace leakage errors), potentially addressing issues of fault-tolerance.
Importantly, these improvements to our implementation may be pursued while preserving the framework of the teleported gate protocol.

This result is not only the first demonstration of a gate between logical qubits, but also the first demonstration of a deterministic teleported two-qubit gate.
A compelling advantage of our work is that the teleported gate is itself modular and uses relatively modest elements, all of which are part of the standard toolbox necessary for quantum computation in general.
Therefore, ongoing progress to improve any of the elements will directly increase gate performance.
Furthermore, the teleported \CNOT protocol used in this work is but one example of an extensive family of two-qubit operations that may be implemented using these same resources \cite{Gottesman1999,Eisert2000,Jiang2007}.
Such teleportation-based gates are important primitives for the implementation of a modular architecture and may be part of a broader approach to fault-tolerant quantum computation \cite{Gottesman1999, Nickerson2013, Monroe2014}.
The next step will be to demonstrate nonlocal teleported gates between spatially separate modules, requiring remote entanglement.
Building on our results and recent demonstrations of remote entanglement in cQED systems \cite{Roch2014,Narla2016}, it should be possible to integrate these technologies in the future.


\begin{acknowledgments}
	We thank B. J. Lester, Z. K. Minev, A. Narla, U. Vool, and I.L. Chuang for fruitful discussions on the manuscript, and A. Narla, K. Sliwa, and N. Frattini for assistance on the parametric amplifier.
	Facilities use was supported by the Yale SEAS cleanroom, YINQE, and NSF MRSEC DMR-1119826.
	This research was supported by the Army Research Office under Grant No. W911NF-14-1-0011 and by Office for Naval Research under Grant No. FA9550-14-1-0052. 
	C.J.A. acknowledges support from the NSF Graduate Research Fellowship under Grant No. DGE-1122492. 
	Y.Y.G. was supported by an A*STAR NSS Fellowship.
	L.J. acknowledges additional support from the Alfred P. Sloan Foundation under Grant No. BR2013-049 and from the Packard Foundation under Grant No. 2013-39273.
	R.J.S., M.H.D., and L.F. are founders, and R.J.S. and L.F. are equity shareholders of Quantum Circuits, Inc.
\end{acknowledgments}

\section*{Contributions}
K.S.C, J.Z.B, and C.S.W. performed the experiment and analyzed the data under the supervision of R.J.S.
P.C.R. developed the feedforward control software and implemented the software used to generate optimal control pulses.
C.J.A., Y.Y.G., and L.F. fabricated the transmon qubits.
K.S.C., J.Z.B., and R.J.S. designed the experiment.
L.J., M.H.D., and L.F. provided theoretical support.
K.S.C. and R.J.S. wrote the manuscript with contributions from all authors.


\newpage

\renewcommand{\theequation}{S\arabic{equation}}
\setcounter{equation}{0}
\renewcommand{\thesection}{S\arabic{section}}
\setcounter{section}{0}
\renewcommand{\thefigure}{S\arabic{figure}}
\setcounter{figure}{0}
\renewcommand{\thetable}{S\arabic{table}}
\setcounter{table}{0}
	
\begin{singlespace}
	
%
%

	\begin{center}
	\MakeUppercase{\textbf{Supplementary Information}}
	\end{center}
	
	
	\section{Experimental Methods\label{sec:intro}}
	
	\subsection{Device}
	
\begin{figure*}[tp]
		\centering{
			\phantomsubcaption\label{subfig:device_full_package}
			\phantomsubcaption\label{subfig:device_top_view}
			\phantomsubcaption\label{subfig:device_sapphire}
		}
		\includegraphics{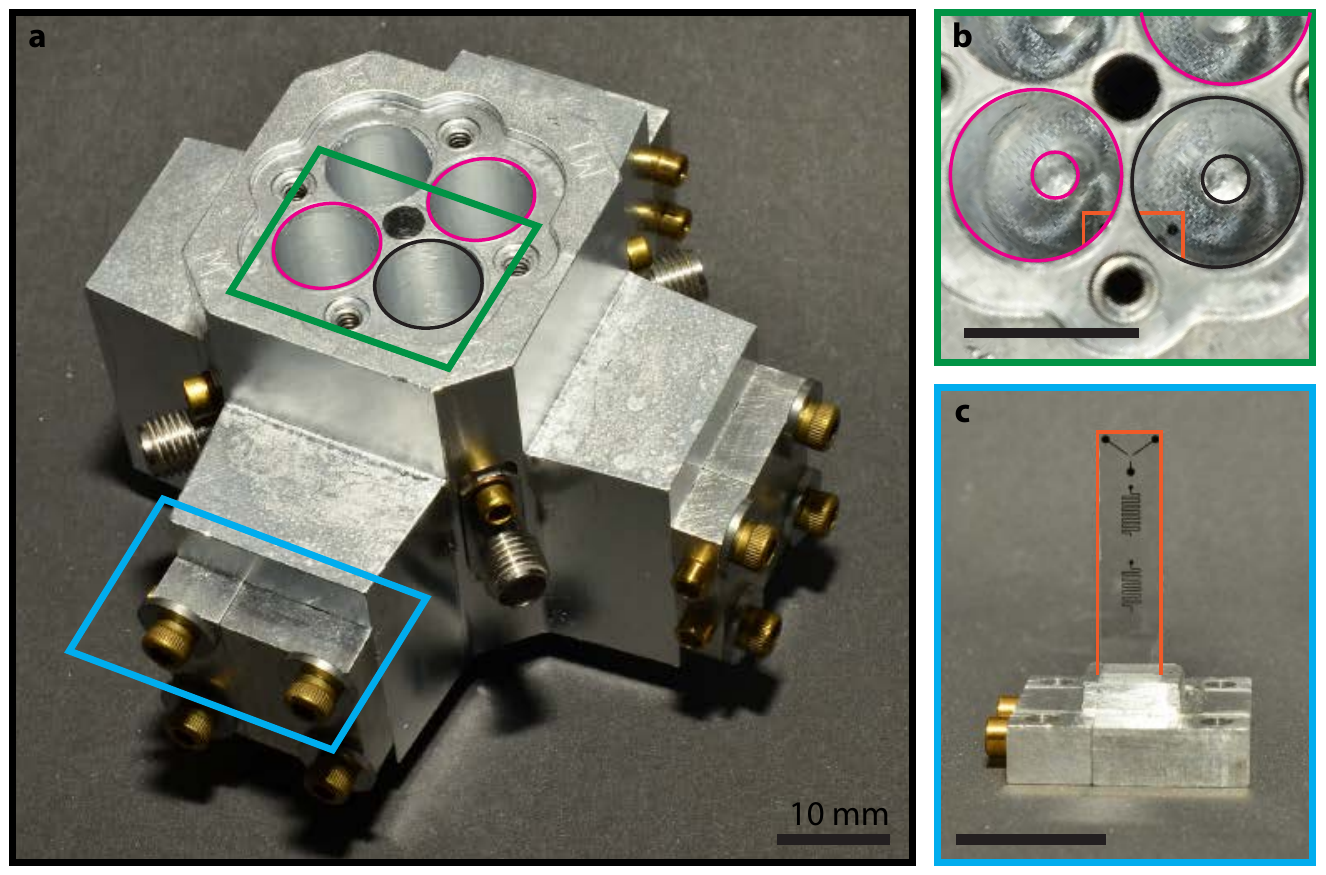}
		\caption{ \label{fig:device_picture}
			\textbf{Overview of physical device.}
			\subref{subfig:device_full_package} Photograph of full device assembly.
			The machined Al package contains four coaxial $\lambda / 4$ 3D cavities, three of which are used in this work. The cavities that serve as data qubits and bus are outlined in pink and black, respectively. A detailed photograph of the cavities is shown in \subref{subfig:device_top_view}. Two clamps anchor each sapphire chip, one is highlighted in blue and is detailed in \subref{subfig:device_sapphire}. The visible connectors are input ports for each cavity; the input/output ports for the transmon and readout resonators are on the underside of the device and thus not visible.
			\subref{subfig:device_top_view} Top-down photograph of cavities.
			We illustrate the three cavities using the same color scheme in \subref{subfig:device_full_package}; the inner circle represents the inner conductor the defines the cavity mode. The orange outline shows the sapphire chip inserted into the device package. Also visible are the antenna pads of the transmon that enable coupling to each cavity.
			\subref{subfig:device_sapphire} Photograph of sapphire chip on which the transmon and readout resonators are fabricated.
			The sapphire chip is outlined in orange and contains several elements: from the top of the figure moving down, the Y-shaped transmon qubit, the readout resonator, and the Purcell filter.
		}
	\end{figure*}

	Our system consists of three 3D cavities, two Y-shaped transmon superconducting qubits, and two Purcell-filtered readout resonator. 
	Our device (\autoref{fig:device_picture}) is constructed from a single machined block of high-purity (4N) Aluminum that physically forms the three 3D cavities as well as the package that houses sapphire chips on which the transmon qubits and quasi-planar readout resonators are defined.
	Each cavity is constructed as a 3D $\lambda / 4$-transmission line resonator, with a cylindrical outer conductor of diameter \SI{9.5}{mm} and stub with inner conductor of diameter \SI{3.2}{mm}. The bottom of the stub transitions into the Al block, thus electrically connecting to the outer conductor and establishing a ground termination. The other end is terminated as an open connection and transforms into a vacuum cylindrical waveguide. These two boundary conditions establish the $\lambda / 4$ resonant structure for the cavity mode. The stub lengths dictate the resonance frequency, and the two data qubit cavities have center pin lengths of \SI{12.7}{mm} and \SI{13.2}{mm}, while the bus cavity has a stub length of \SI{11.7}{mm}. 
	The far-end of all three cylindrical waveguides are closed off with a separate Al cap. This waveguide, physically necessary for machining the cavity stub, also serves to isolate the cavity electromagnetic mode from a potentially lossy seam formed by the opening at the top of the waveguide. Based on the cutoff frequency of \SI{20}{GHz}, the seam energy participation in the cavity mode is $<10^{-8}$. 
	For the transmon qubits to gain access to the cavities, tunnels are opened up in the Al block perpendicular to the axial axis of the cavities. We position two such tunnels so that each individually intersects with one data qubit cavity and the bus cavity. The tunnels are located at a height near the top of the stubs to maximize the electric field coupling between the cavity and transmon qubit.
	Into these tunnels, we insert a sapphire chip onto which we have lithographically printed the transmon qubit and readout resonator.
	The entire machined Al package is chemically etched around \SI{100}{\us} to improve the surface quality by removing machining damage \cite{Reagor2013}.
	
	On the sapphire chip, the transmon qubit is designed with Y-shaped antenna pads \cite{Wang2016} to couple to three distinct modes: the cavity that encodes the data qubit, the bus, and the readout resonator. When the chip is inserted in the tunnel, the two arms of the antenna protrude into the space of data qubit and bus cavity. This enables capacitive coupling between each cavity and the transmon qubit (and to a lesser degree, mode mixing between the two 3D cavities). On the opposite side of these antenna pads, we print two strips of Al that form the centerpin of quasi-planar $\lambda / 2$ stripline resonators. One functions as the readout resonator mode and the other functions as band-pass, Purcell filter \cite{Axline2016} to protect the transmon qubit and 3D cavities from spontaneous emission into the readout resonator. The striplines are constructed with meanders to decrease the physical footprint of each resonator while maintaining a particular mode frequency.
	The readout resonator (and filter mode) is strongly coupled to a $50 \Omega$ transmission line for fast readout of the transmon qubit state.
	We use a standard electron-beam lithography process to simultaneously define the transmon qubits and stripline resonators.
	Our transmon qubit Josephson junctions are defined with the Bridge-free shadow mask process \cite{Lecocq2011}.
	Each chip is diced from a wafer of \SI{430}{\um} c-plane sapphire to dimensions of $\SI{5.5}{mm} \times \SI{27.5}{mm}$.
	
	\subsection{System Hamiltonian} \label{sec:hamiltonian}
	
	The system of nine modes can be split up into two categories: nearly-linear harmonic oscillator modes (describing the cavities, resonators, and filters) and anharmonic bosonic modes (describing the transmon qubits). In this work we utilize the lowest two levels of the transmon qubits.
	A detailed table of coherences in our experiment is provided in \autoref{table:coherences}.
	We use the finite-element 3D simulation package \textsc{ANSYS}\textregistered\xspace HFSS \cite{HFSS} and Black-box quantization \cite{Nigg2012} to transform the physical 3D architecture described above into the dispersive Hamiltonian model.
	
	\begin{table*}[tp] 
		\begin{tabular}{c c c c c }
			\hline \hline
			module	& 	mode 	& Energy relaxation time, & Ramsey dephasing time, & Echo dephasing time, \\ 
			& 			& $\tau_{1ph}$ or $T_1$ ($\si{\us}$)
			& $T_{2}^{\text{R}}$ ($\si{\us}$) & $T_{2}^{\text{E}}$ ($\si{\us}$) \\
			\hline \hline
			1 	& data 			& $1150$ 	& $390$ 	& -- \\
			& communication 	& $65 - 69$ 		& $11 - 14$ & $18-20$\\
			& readout			& $0.1$ 	& -- 		& -- \\
			\hline
			2 	& data 			& $1100$ 	& $390$ 	& -- \\
			& communication 	& $67 - 77$ 		& $18 - 22$ & $22 - 24$\\
			& readout			& $0.1$ 	& -- 		& --\\
			\hline
			& bus 				& $230$ 	& -- 		& -- \\
			\hline \hline
		\end{tabular}
		\caption{\label{table:coherences}
			\textbf{System coherences.}
		}
	\end{table*}
	
	To understand our system Hamiltonian we first define a few primitives.
	\begin{itemize}
		\item Kerr oscillator Hamiltonian, which describes an anharmonic oscillator
		\begin{equation} \label{eq:H_anharmonic_osc}
		\hat{H}_{O}(\hat{a}) = \omega_a \hat{a}^\dagger \hat{a} - \frac{K_a}{2} \hat{a}^\dagger \hat{a}^\dagger \hat{a} \hat{a},
		\end{equation}
		where $\omega_a$ represents the resonance frequency and $K_a$ is the self-Kerr, or anharmonicity, for mode $\hat{a}$.
		\item Dispersive coupling Hamiltonian, which describes the dispersive interaction between two modes $\hat{a}$ and $\hat{b}$:
		\begin{equation} \label{eq:H_disp}
		\hat{H}_{\textrm{disp}}(\hat{a}, \hat{b}) = 
		\underbrace{-\chi_{ab} \hat{a}^\dagger \hat{a} \hat{b}^\dagger \hat{b}}_{\hat{H}^{(0)}_{\textrm{disp}}(\hat{a}, \hat{b})}
		\underbrace{+\chi'_{ab} (\hat{a}^\dagger)^2 (\hat{a})^2 \hat{b}^\dagger \hat{b}}_{\hat{H}^{(1)}_{\textrm{disp}}(\hat{a}, \hat{b})},
		\end{equation}
		where $\chi_{ab}$ is the dispersive interaction between modes $a$ and $b$ and $\chi'_{ab}$ is the nonlinear dispersive interaction (i.e. an interaction dependent on the number of photons in each mode). Typically we take $\hat{b}$ as the transmon mode, and since we will operate in the two-level subspace, we can safely ignore the other nonlinear term, $\hat{a}^\dagger \hat{a} (\hat{b}^\dagger)^2 (\hat{b})^2$, can be ignored. Otherwise, we will be explicit if we only consider one of the dispersive terms.
	\end{itemize}
	
	We can then write the Hamiltonian for each module, which includes one 3D cavity with operator, $\hat{c}$; one transmon qubit, $\hat{q}$; and one readout resonator, $\hat{r}$ (the filter resonator is not included as it is never directly populated and thus, does not participate in the Hamiltonian dynamics of the system).
	\begin{equation} \label{eq:Hamiltonian_module}
	\begin{aligned}
	\hat{H}_{\textrm{module}}(\hat{c}, \hat{q}, \hat{r}) &= \hat{H}_O(\hat{c}) + \hat{H}_O(\hat{q}) + \hat{H}_O(\hat{r}) \\
	&+ \hat{H}_{\textrm{disp}}(\hat{c}, \hat{q}) + \hat{H}_{\textrm{disp}}(\hat{r}, \hat{q}) \\
	&+ \hat{H}^{(0)}_{\textrm{disp}}(\hat{c}, \hat{r}).
	\end{aligned}
	\end{equation}
	The relevant parameters includes the mode frequencies ($\omega_c$, $\omega_q$, $\omega_r$); 
	the self-Kerrs ($K_c$, $K_q$, $K_r$); the dispersive interaction between the readout resonator and transmon ($\chi_{rq}$ and $\chi'_{rq}$); 
	and finally, the cross-Kerr between the cavity and readout resonator ($\chi_{cr}$). 
	In practice, we neglect a few terms: first, the self-Kerr of readout resonator, $K_r$, which typically causes a small perturbation on the readout resonator response during transmon measurement; and second, the nonlinear interaction term between cavity and readout resonator, $\chi'_{cr}$, is small as both modes are nearly-linear harmonic oscillators and can be ignored. 
	Hamiltonian parameters for each module are tabulated in \autoref{table:hamiltonian_paramaters}.
	
	Next, we group the main terms that participate in the dynamics for entangling the two transmon (communication) qubits: the two transmon qubits ($\hat{q}_1$ and $\hat{q}_2$) and bus cavity ($\hat{b}$).
	This Hamiltonian is described as follows:
	\begin{equation}
	\begin{aligned}
	\hat{H}_{\textrm{coupling}}(\hat{q_1}, \hat{q_2}, \hat{b}) 
	&= \hat{H}_O(\hat{q}_1) + \hat{H}_O(\hat{q}_2) + \hat{H}_O(\hat{b}) \\
	&+ \hat{H}_{\textrm{disp}}(\hat{b}, \hat{q}_1) + \hat{H}_{\textrm{disp}}(\hat{b}, \hat{q}_2) \\
	&+ \hat{H}_{\textrm{disp}}(\hat{q}_1, \hat{q}_2),
	\end{aligned}
	\end{equation}
	The relevant parameters includes the mode frequencies ($\omega_{q_1}$, $\omega_{q_2}$, $\omega_b$) and the self-Kerrs ($K_{q_1}$, $K_{q_2}$, $K_r$) as well as the interaction terms between the bus and each qubit, $\chi_{bq_{1,2}}$ and $\chi'_{bq_{1,2}}$. Importantly, we also include the direct interaction between the two transmon qubits, $\chi_{q_{1}q_{2}}$. This term, as we will see, is the dominant residual coupling between the two modules. As we operate in the two-level subspace of both transmon qubits, the nonlinear interaction terms, $\chi'_{q_{1,2}q_{2,1}}$, do not play a role in the dynamics of our experiment.
	Hamiltonian parameters for this subsystem are tabulated in \autoref{table:hamiltonian_paramaters}.

	\begin{table*}[tp] 
		\begin{ruledtabular}
			\begin{tabular}{c c c c c c}
				& 		& 						& \multicolumn{3}{c}{Mode coupling, $\chi_{ij} / (2 \pi)$ (MHz)} \\
				\noalign{\vskip 1pt} \cline{4-6} \noalign{\vskip 1pt}
				Subsystem	& Mode			& Mode frequency	& Data qubit	&  Communication qubit	& Readout \\
				\hline \noalign{\vskip 1pt}
				Module 1	& Data qubit			& 5123.6	& \scinotation{1.1e-3} 	& $0.573 + 0.00061 \hat{n}$ & $\le10^{-3}$  \\
				& Communication qubit	& 4387.7	& ---			& 131.2	& 2.7 	\\
				& Readout resonator		& 7720.0	& ---			& ---	& --- \\
				\noalign{\vskip 1pt} \hline \noalign{\vskip 1pt}
				Module 2	& Data qubit			& 5275.0	& \scinotation{1.8e-3} 	& $0.843 + 0.0014 \hat{n}$ & $\le10^{-3}$  \\
				& Communication qubit	& 4559.2	& ---			& 123.2	& 2.8 \\
				& Readout resonator		& 7735.4	& ---			& ---	& --- \\
				\hline \noalign{\vskip 6pt} \hline \noalign{\vskip 1pt}
				Subsystem	& Mode			& Mode frequency	& Bus cavity	&  Communication qubit 1	& Communication qubit 2 \\
				\hline \noalign{\vskip 1pt}
				Coupler		& Bus cavity			& 5692.8	& \scinotation{0.3e-3} 	& $0.319 + 0.001 \hat{n}$ &  $0.455 + 0.001 \hat{n}$  \\
				& Communication qubit 1	& 4387.7	& ---			& ---	& 0.019 \\
				& Communication qubit 2	& 4559.2	& ---			& ---	& --- \\
			\end{tabular}
		\end{ruledtabular}
		\caption{\label{table:hamiltonian_paramaters}
			\textbf{Measured Hamiltonian parameters.}
		}
	\end{table*}
	
	\subsection{Validity of module approximation}
	A defining characteristic of modular architectures---which makes it advantageous for scaling quantum systems---is that data qubits are housed in separate modules to ensure that residual interactions can be made vanishingly small.
	Here, the modules in our experiment are coupled via local means, and we assess the validity of treating them separately by discussing the magnitude of residual interactions in our system. 
	Our implementation requires purely local (capacitive) coupling to build the Hamiltonians discussed above and, if uncontrolled, some of these interactions can extend to disparate parts of the system.
	
	The dominant residual interaction in our system is $\chi_{q_{1}q_{2}}$, the direct coupling between the two transmon qubits. This term arises from a combination of two contributions \cite{Dalmonte2015}: a direct dipole-like coupling and a cavity-mediated interaction. In our system, the this term arises primarily from the mutual interaction with the cavity mode. This term remains a perturbation on our experiment for three reasons: first, the magnitude of this interaction is over an order of magnitude smaller than other relevant terms in the Hamiltonian; second, we choose the anti-symmetric, single-excitation Bell pair for the communication (transmon) qubits where this term does not statically participate; third, we design our local operations to be insensitive to a small frequency transmon qubit frequency shift.
	
	Next, we consider the cross-Kerr interactions between the data qubit and the bus: $\chi_{c_{i} b}$. Experimentally, we have bounded this term to be less than a few $\si{kHz}$ via Stark shift measurements, consistent with a fourth order approximation, where we estimate the interaction to be, $\chi_{cb} \approx \chi_{qb} \chi_{qc} / (2 K_q) \approx 1 \textrm{kHz}$, where mode $q$ represents the transmon qubit that couples to both data qubit and bus. This interaction term is never directly involved during the teleported gate protocol. First, the bus cavity only is manipulated during the communication qubit Bell state generation step and otherwise it is left in the vacuum state where this term does not contribute to any system dynamics. Even during the Bell state generation, the bus is driven off-resonantly and the bus is never directly populated.
	
	The final residual interaction and most critical to our approximation is the direct coupling between the two data qubits $\chi_{c_1 c_2}$. This term describes the rate at which the two cavities naturally entangle and is a metric with which we compare the interaction time for our teleported gate. We have not extracted statistically significant measurements of this residual interaction term. Indeed, from the simulated Hamiltonian from black box quantization, we roughly expect couplings on the order of $\SIrange[range-phrase = -, range-units=single]{1}{10}{\Hz}$, an immeasurably small quantity in our system compared to all other interaction strengths. Indeed, this bound on the coupling rate is at least an order of magnitude smaller than the data qubit decay rate, $\kappa / 2\pi \approx \SI{160}{Hz}$. Given this analysis, we conclude that for the purpose of our work, the two data qubits are non-interacting despite the local nature of our device package.
	
	Though these simplifications are reasonable, we note that in a future implementation with remote logical qubit modules, these residual interactions will be completely obviated.
	
	\subsection{Experimental setup}
	
	\subsubsection{Cryogenic hardware}
\begin{figure*}[tp]
		\centering
		\includegraphics[width=170mm]{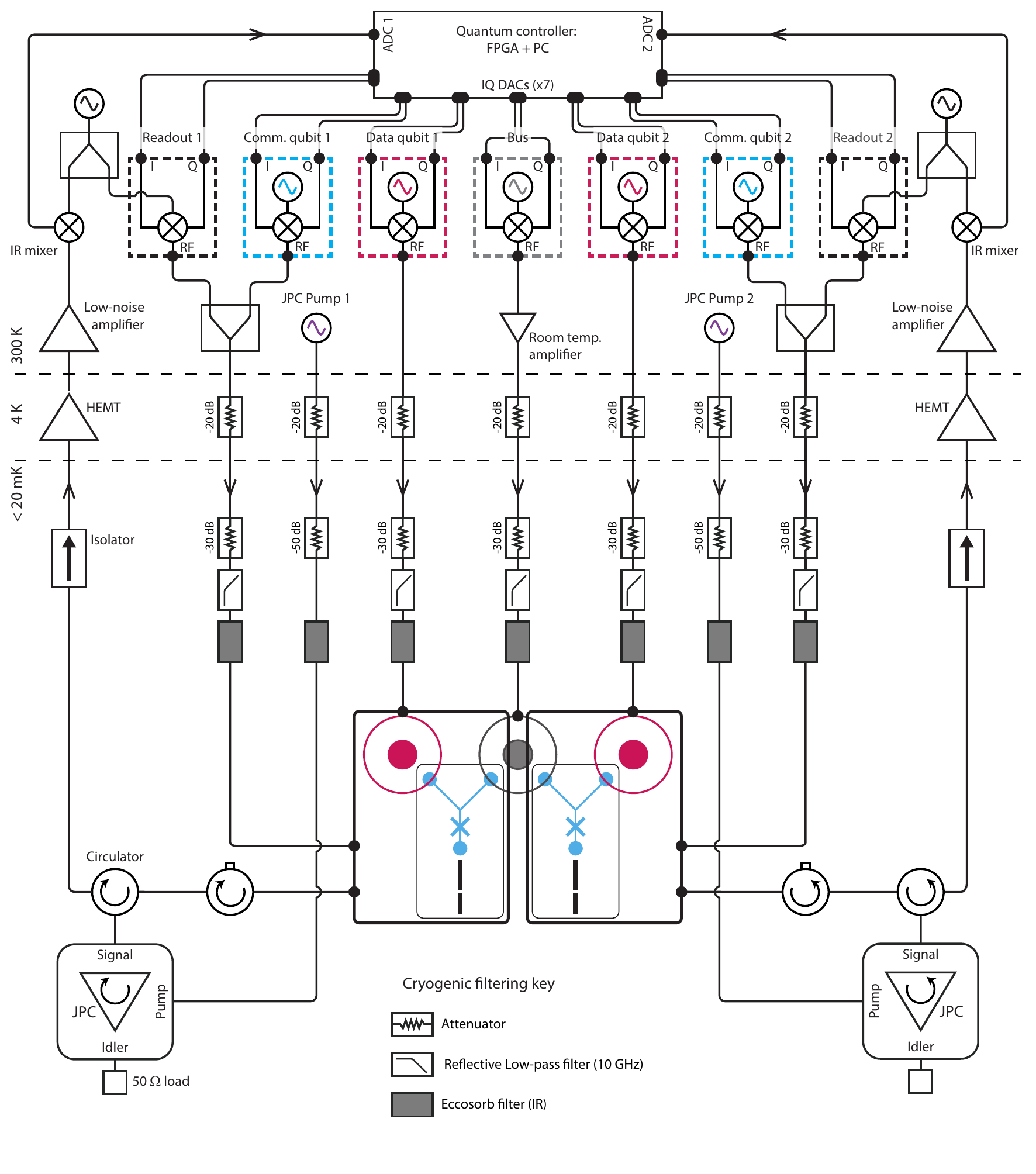}
		\caption{\label{fig:wiring_diagram}{
				\textbf{Detailed experimental setup}.
				Input control signals are generated at room temperature using standard cQED microwave techniques and are attenuated and filtered in the refrigerator before reaching the device.
				At room temperature, each signal is generated using a combination of an RF-microwave generator (data and communication qubits: \emph{Vaunix LabBrick LMS-103-13}, readout resonators and bus: \emph{Agilent/Keysight N5183A} and \emph{E8275D}) and IQ mixer setup (\emph{Marki Microwave IQ-0307LXP}).
				The bus drive line also includes an amplifier (\emph{MiniCircuits ZVA-183-S+}) and fast microwave switch (\emph{Hittite HMC-C019}) at room temperature. 
				A custom quantum control computer (\emph{Innovative Integration VPXI-ePC with four X6-1000M boards}) calculates and generates IF signals in real-time.
				Output signals eminate from the strongly coupled port of a readout reasonator and travel through two circulators (\emph{Quinstar}), are amplified by a Josephson parametric converter (JPC) that is continuously pumped by microwave generator (\emph{Agilent/Keysight N5183A}).
				These signals then travel through superconducting transmission lines to an additional cryogenic (\emph{Low Noise Factory LNF-LNC4\_8C}) and room-temperature amplifiers (\emph{Miteq AFS3-00101200-35-ULN}) before being mixed down (\emph{Marki Microwave IR-4509}) to be demodulated and analyzed by the control computer.
		}}
	\end{figure*}

	The entire package is anchored through a Oxygen Free High Conductivity copper bracket to the mixing chamber stage of a dilution refrigerator which is cooled down to 10 - 20 mK. To further isolate the device from residual magnetic fields a Cryoperm magnetic shield is mounted around the package. Inside of the magnetic shield we include a copper sheet with coated with Stycast and Carbon Black to absorb stray radiation that is inside of the magnetic shield. Microwave attenuators, low-pass filters, and absorptive, infrared (Eccosorb\textregistered) filters are used to reduce radiation and noise on each control microwave input line. We use two Josephson Parametric Converters (JPC) in this experiment, and these are housed in separate bracket-shield configurations on the mixing chamber. Each JPC is connected to an output port on our device through a pair of circulators to ensure one-way signal propogation from the device to the JPC. Both JPCs provide nearly quantum-limited amplification of around 20 dB at a bandwidth of around \SI{5}{\MHz} and a Noise Visibility Ratio \cite{Narla2014} of \SI{4}{dB}. Specific wiring details can be found in \autoref{fig:wiring_diagram}.
	
	\subsubsection{Microwave control}
	All modes in the system are controlled at room temperature using microwave-frequency pulses generated through single-sideband modulation (SSB) of an IQ mixer. For each input, a dedicated microwave generator serves as a local oscillator that up-converts shaped intermediate-frequency (IF) control pulses generated by our quantum controller. We utilize an all-in-one quantum control architecture first introduced in \cite{Ofek2016} that combines three requirements necessary for performing quantum experiments into one control system: generate and output pulses to manipulate the quantum system, sample input signals for quantum measurements, and crucially, perform classical computation based on these input signals to determine the next output. These three requirements form a closed cycle of quantum control and enable the deterministic implementation of the teleported gate as presented in this work. Our controller accomplishes all of these tasks on the same timescale in which our experiments operate on, enabling real-time control, either feedback or feedforward, on individual experimental shots.
	
	Each mode (excluding the filter resonators) is controlled with a dedicated IQ mixer setup, enabling fast and individual control. In particular, we drive operations on the transmon $\ket{g} \leftrightarrow \ket{e}$ levels as well as linear displacements on the cavity modes using Gaussian shaped pulses with $\sigma = 6$ ns (and total extent of $4\sigma$). 
	We also apply derivative removal via adiabatic gate (DRAG) to account for the presence of higher levels of the transmon qubits, and any residual state leakage or phase error is a small source of infidelity in our experiment.
	Off-resonant pump drives are used in several cases for this work and these are generated using the same physical hardware; our controller is able to generate multiple sideband frequencies on a single IF digital-to-analog output.
	
	\subsubsection{Transmon measurement}
	In this experiment, each module is connected to a separate JPC for fast, high-fidelity measurement of the transmon qubit. We achieve single-shot assignment fidelities around $99.4\%$, largely limited by transmon decay during the measurement pulse of 600 ns. We define assignment fidelity as the average of probabilities of correctly assigning the state when we prepare the transmon in $\ket{g}$ and $\ket{e}$: $F_{assign} = \left[\Pr(\text{``g''}|\ket{g}) + \Pr(\text{``e''}|\ket{e})\right]/2$. This high quality measurement, coupled with the real-time capabilities of our quantum controller, enable classically conditional operations based on an extracted measurement result. The length of time from the start of a measurement pulse to the application of a conditioned operation is around \SI{1000}{ns}, which includes measurement pulse length (\SI{600}{ns}), cable delays (\SI{200}{ns}), integration and state estimation latencies (\SI{200}{ns}). 
	
\begin{figure*}[tp]
		\centering{
			\phantomsubcaption\label{subfig:meas_crosstalk_circuit}
			\phantomsubcaption\label{subfig:meas_crosstalk_data_1}
			\phantomsubcaption\label{subfig:meas_crosstalk_data_2}
		}
		\includegraphics{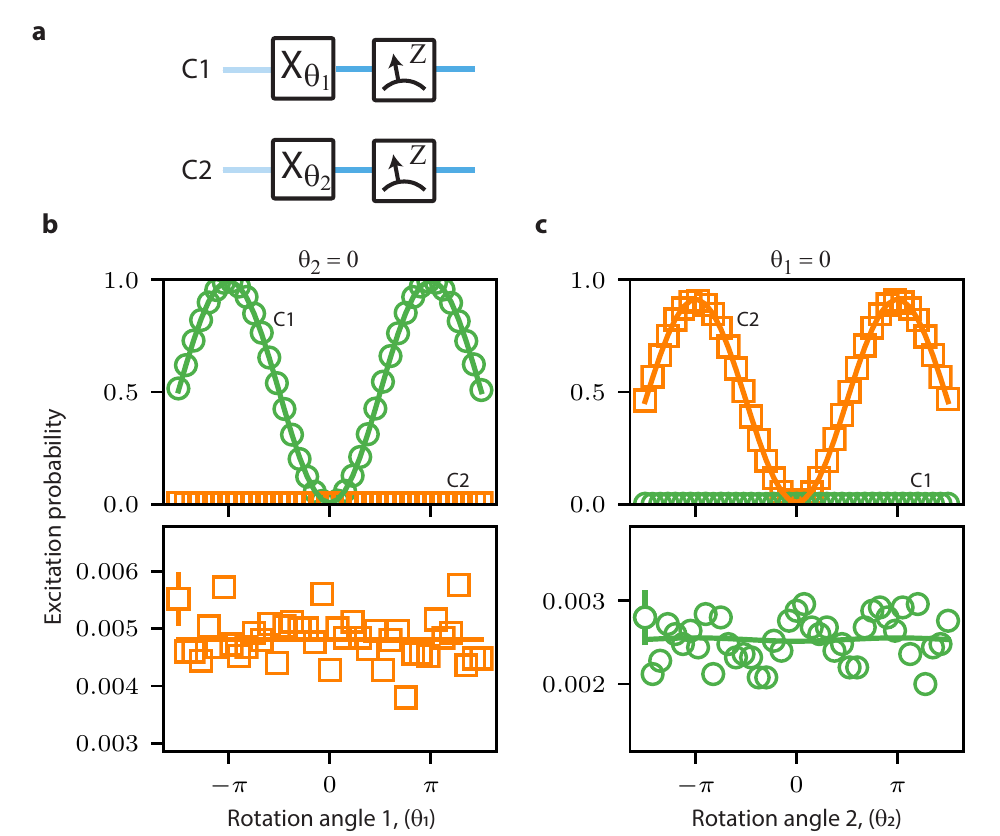}
		\caption{ \label{fig:meas_crosstalk}
			\textbf{Assessing independence of single qubit measurements.}
			\subref{subfig:meas_crosstalk_circuit} Rabi experiment pulse sequence to extract measurement crosstalk.
			After initializing both communication qubits in the ground state, both qubits are rotated by $\hat{X}$-rotations with independent angles $\theta_1$ and $\theta_2$ for C1 and C2, respectively. Subsequently, measurements are performed on modules 1 and 2 and the result is recorded.
			\subref{subfig:meas_crosstalk_data_1} and \subref{subfig:meas_crosstalk_data_2} Measurement crosstalk experimental results. 
			For \subref{subfig:meas_crosstalk_data_1} (\subref{subfig:meas_crosstalk_data_2}), C2 (C1) is kept in the ground state, and a Rabi experiment is performed on C1 (C2). 
			The measurement results are shown for C1 (green circles) and C2 (orange squares).
			For clarity, we describe the results focusing on \subref{subfig:meas_crosstalk_data_1}; the discussion is the same for \subref{subfig:meas_crosstalk_data_2}, save swapping C1 and C2.
			Top panel: the C1 measurement results illustrate high contrast oscillations, while the C2 measurement results remains close to zero, as expected when the communication qubit measurements are independent.
			Bottom panel: Zoom in for measurement results on C2. The lack of structure in the data indicate that the measurement of C2 does not infer any information about the state of C1.
			To estimate the measurement crosstalk, we perform sinusoidal fits the data by fixing the frequency and phase of the oscillation and extracting an amplitude and offset. 
			Each point in this experiment consists of $25,000$ experiments. For data in the top panels, error bars are much smaller than the marker; for data in the bottom panels, we represent a typical error bar to be within the spread of the points.
			The slightly reduced contrast in \subref{subfig:meas_crosstalk_data_2} is specific to this calibration experiment, potentially due to drifts of transmon relaxation rate during the many hours of acquisition.
		}
	\end{figure*}

	It is critical that the two communication qubit measurements be independent for the demonstration of the teleported gate. 
	In order to assess the measurement crosstalk, we perform a Rabi experiment and perform simultaneous measurements on both communication qubits (\autoref{subfig:meas_crosstalk_circuit}). Our results \autoref{subfig:meas_crosstalk_data_1} and \autoref{subfig:meas_crosstalk_data_2} indicate that the measurements are highly selective to the qubit addressed. 
	From our data, we estimate the measurement crosstalk--defined to be the ratio of the measurement contrast of measuring the directly coupled qubit to that of measuring the isolated qubit--to be $< 10^{-4}$.
	In future implementations of this experiment where the two modules are physically separate, the measurement crosstalk will be completely neglegible.
	

	\subsubsection{System preparation} \label{sec:system_prep}
	Ideally, with the device thermalized to the mixing chamber of the dilution refrigerator at \SI{20}{mK}, all modes should be in ground state at equilibrium. However, in our experiment and common to many cQED experiments, we find that each mode has a non-negligible thermal population, potentially due to improper thermalization or additional input/output line noise. In particular, the equilibrium population of our transmon qubits and cavities are around $10\%$ and $< 1\%$, respectively. To ensure accurate state preparation for our protocol, it is necessary to address this residual population, and we do so by performing a full-system feedback cooling protocol at the start of every experiment. With this cooling protocol, the entire system is prepared in the ground state (transmons in the $\ket{g}$ state and cavities in $\ket{0}$) with fidelity in excess of $99\%$.
	
\begin{figure*}[tp]
		\centering{
			\phantomsubcaption\label{subfig:fb_cooling_transmon}
			\phantomsubcaption\label{subfig:fb_cooling_full}
		}
		\includegraphics{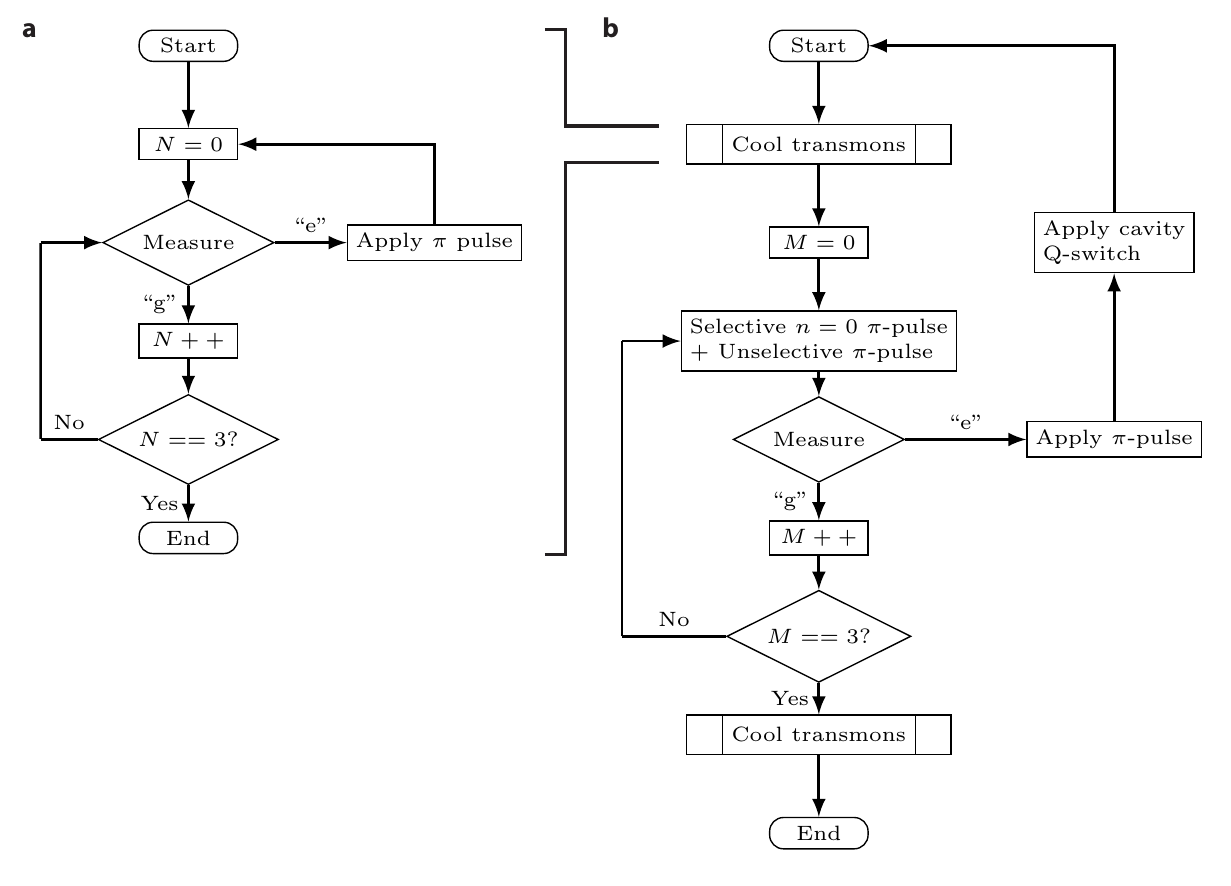}
		\caption{ \label{fig:fb_cooling}
			\textbf{Cooling sequences.}
			Measurement-based feedback sequences used for system-wide ground state preparation.
			\subref{subfig:fb_cooling_transmon} Cooling sequence used to reset both of the transmon (communication) qubits. 
			To increase confidence in the ground state preparation, we require three consecutive measurements that both qubits are in the ground state before accepting the reset has succeeded.
			\subref{subfig:fb_cooling_full} Cooling sequence used to reset the entire system.
			The transmon cooling sequence is used as a subroutine for the full system reset.
			We apply a long selective $\pi$ pulse on each transmon that flips the transmon qubit only when the data qubit and bus are both in the vacuum state. An additional unselective $\pi$-pulse so that the successful measurement outcome indicates that the transmon qubit is in the ground state. 
		}
	\end{figure*}

	Our feedback cooling protocol is diagrammed in \autoref{fig:fb_cooling}. It consists of three parts: (1) reset the transmon qubits to $\ket{g}$, (2) empty the cavities to the vacuum state, (3) check that the transmon qubits are still in the ground state. In Step 1, the transmon qubit is first measured; if the result indicates that the qubit is in the excited state $\ket{e}$, then the controller dynamically applies a $\pi$ pulse to flip the state down to $\ket{g}$. This process is repeated several times to build confidence that the transmon is in $\ket{g}$, continuing to the next step only if there have been three consecutive ``g'' measurement results. In our experiment, we cool both transmon qubits simultaneously, and so the signature for successful measurement is the joint ground state $\ket{gg}$. In Step 2, to cool the cavities, we apply simultaneous $\pi$ pulses on the two transmon qubits that flips the each qubit when and only when their respective data qubit cavities are in the vacuum state $\ket{0}$. We use long $\sigma = 1 \mu s > 1 / \min{\chi}$ to ensure that we are selective on the transmon resonance frequency when the cavity is in $\ket{0}$. This protocol also provides a check that the bus cavity is in the vacuum state as well. These selective pulses have a duration that is an appreciable fraction of transmon coherence times, and thus have a lower probability of correctly flipping the qubit ($\sim 90\%$). To account for this diminished contrast, we repeat this process three times, only continuing when we have three consecutive successful measurements. If a failure occurs indicating that the data qubit cavities have some finite population, then we apply a four-wave mixing process (Q-switch) that rapidly evacuates population in the data qubit cavities into the readout resonator with a time constant of $\tau \sim \SI{100}{\us}$ \cite{Pfaff2017}. We then return to Step 1. When we successfully complete Step 2, we then perform a feedback cooling check similar to Step 1 to ensure that the transmon qubits are in the ground state. This last step typically takes less than $\SI{10}{\us}$, which is far shorter compared to the estimated heating rate of the data qubit cavities of $\SIrange[range-phrase = -, range-units=single]{10}{100}{ms}$. This cooling protocol is efficient, enabling an experimental repetition period of around \SI{1}{\us}.
	
	\section{Teleported \CNOT gate protocol}
	
	\subsection{Implementation}
	
	Here, we provide a more detailed description of our implementation of the teleported \CNOT gate. 
	In \autoref{sec:toolbox}, we describe the experimental toolbox we use for this experiment.
	In \autoref{sec:tuneup}, we provide additional information regarding the tuneup protocol of the teleported \CNOT, specifically the critical importance of tracking the data qubit reference frame for this operation.
	In \autoref{subfig:telecnot_circuit}, we show a detailed circuit representation of the teleported \CNOT protocol, and in \autoref{subfig:telecnot_timing}, we illustrate the timing for the entire protocol. 
	
	Our experiment begins with a system-wide ground state preparation using a measurement-based feedback protocol previously discussed in \autoref{sec:system_prep}.
	After all components of the system are initialized in the ground state, a chosen initial state is encoded onto the data qubits (\autoref{sec:data_encodings}).
	This is accomplished by generating the initial state in the communication qubits and then applying an encoding optimal control pulse that transfers this state onto the logical basis of the data qubits (\autoref{sec:local_ops}). 
	After this encoding step, the communication qubits end in the ground state ready for further use.
	Next, we generate an entangled pair between the communication qubits, and during this operation, the data qubits store the encoded quantum information (\autoref{sec:comm_bell_pair}). 
	Next, we perform the requisite local operations between the data and communication qubits by applying optimal control pulses within each module (\autoref{sec:tuneup_local_ops}).
	Next, both communication qubits are measured in the appropriate basis ($\hat{Z}$ and $\hat{X}$) and subsequently reset to the ground state for reuse (\autoref{sec:comm_meas_reset}).
	The measurement outcomes are distributed and the appropriate feedforward operations are applied to implement a deterministic operation, one that is independent of the measurement outcome.
	Finally, we analyze the output state by performing Wigner tomography or logical quantum state tomography (QST) on the data qubits (\autoref{sec:tomography}).
	Our implementation of Wigner tomography matches the sequences used in \cite{Vlastakis2013}, requiring a Ramsey sequence that maps the photon number parity of the cavity state onto the state of the communication qubit.
	Our implementation of logical QST requires a decoding step, where we use an optimal control pulse to map the encoded data qubit state onto the communication qubits. We then proceed with standard QST by performing the required rotations and measurements directly on the communication qubit.
	
	\subsection{Walkthrough of the teleported gate}
	
\begin{figure*}[tp]
		\centering{
			\phantomsubcaption\label{subfig:telecnot_circuit}
			\phantomsubcaption\label{subfig:telecnot_timing}
		}
		\includegraphics{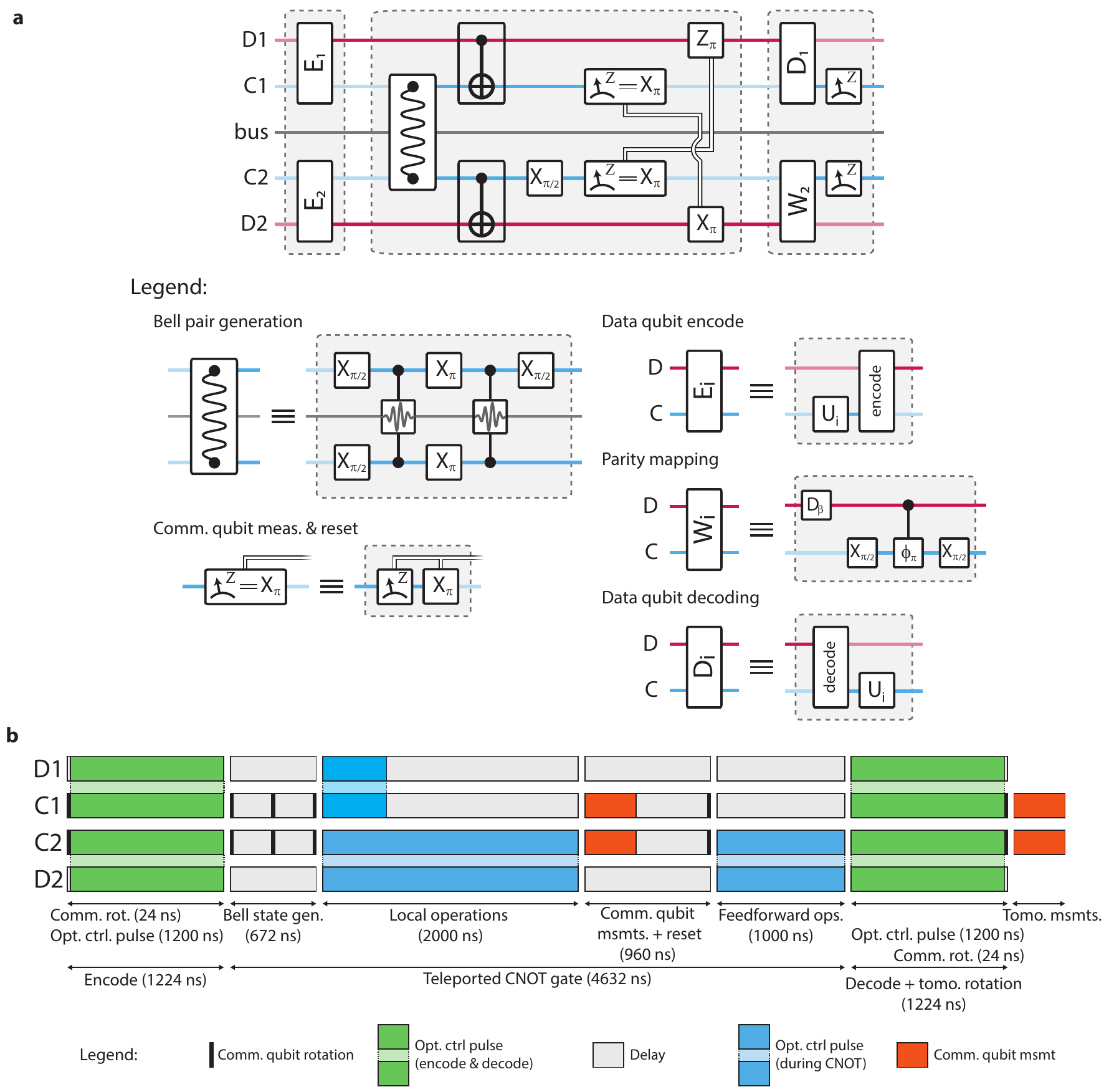}
		\caption{\label{fig:teleported_cnot}{
				\textbf{Teleported \CNOT gate implementation.}
				\subref{subfig:telecnot_circuit} Detailed circuit diagram for the teleported \CNOT gate protocol.
				Top: Pulse sequence for an example experiment. Bottom: Legend for specific circuit blocks.
				In the first panel, we show our sequence for encoding quantum information onto the data qubit. 
				In the second panel, we illustrate our implementation of the teleported \CNOT gate. 
				We show the pulse sequence used to implement the communication qubit Bell state generation.
				For the communication qubit measurements, we apply a $\pi/2$ rotation on C2 in order to measure $\hat{X}$. After the measurement we also perform a measurement-based reset of both C1 and C2 before performing feedforward operations on the data qubits.
				In the third panel, we detail two possible sequences for extracting the data qubit state. 
				For module 1, we perform logical tomography on the data qubits by decoding the data qubit onto the communication qubit and performing the appropriate tomography rotations on the communication qubit.
				For module 2, we perform Wigner tomography by performing a parity mapping sequence on the communication qubit.
				\subref{subfig:telecnot_timing} Teleported \CNOT gate timing diagram.
				The teleported \CNOT is illustrated taking the relative timing of each element into account. The diagram is color-coded with the following designations: single communication qubit rotations in black; encode and decode (optimal control) operations in green; the teleported \CNOT local operations (also optimal control) in blue; and the measurements in orange. This presentation provides a visual representation of the relative durations for each part of the protocol. The teleported \CNOT takes in total around \SI{4.6}{\us}.
		}}
	\end{figure*}

	Here we walk-through the teleported \CNOT gate in \autoref{fig:teleported_cnot}. To better distinguish the data from communication qubit, we will use numerical kets $\lbrace \ket{0}, \ket{1} \rbrace$ for the data qubit and energy-level kets $\lbrace \ket{g}, \ket{e} \rbrace$ for the communication qubit. 
	\begin{enumerate}
		\item We initialize the system in a general two-qubit state for the data qubit $\ket{\psi_{12}}$ and the anti-symmetric Bell state $\ket{\Psi^+} = \left( \ket{ge} + \ket{eg} \right) / \sqrt{2}$ for the two communication qubits.
		\begin{equation}
		\begin{aligned}
		\ket{\psi_{12}} \ket{\Psi^+} &=  
		\left( a \ket{00} + b \ket{01} + c \ket{10} + d \ket{11} \right) \\
		&\otimes \frac{1}{\sqrt{2}} \left( \ket{ge} + \ket{eg} \right),
		\end{aligned}
		\end{equation}
		where the data qubit state is parameterized with four complex probability amplitudes that obey the normalization constraint, $|a|^2 + |b|^2 + |c|^2 + |d|^2 = 1$. For the remainder of the discussion we will incorporate the appropriate normalization into these coefficients.
		\item First, we perform the control-module local \CNOT, a data-qubit controlled, communication-qubit target gate: 
		\begin{equation}
		\begin{aligned}
		\ket{\psi}_{\CNOT_\text{control}} 
		&= a \ket{00ge} + a \ket{00eg} \\
		&+ b \ket{01ge} + b \ket{01eg} \\
		&+ c \ket{10ee} + c \ket{10gg} \\
		&+ d \ket{11ee} + d \ket{11gg} \\
		\end{aligned}
		\end{equation}
		\item Second, we perform the target-module local \CNOT, a communication-qubit controlled, data-qubit target gate: 
		\begin{equation} 		
		\begin{aligned}
		\ket{\psi}_{\CNOT_\text{target}} 
		&= a \ket{01ge} + a \ket{00eg} \\
		&+ b \ket{00ge} + b \ket{01eg} \\
		&+ c \ket{11ee} + c \ket{10gg} \\
		&+ d \ket{10ee} + d \ket{11gg} \\
		\end{aligned}
		\end{equation}
		\item Next, in order to measure the $\hat{X}$-basis on the target-module communication-qubit, we perform a $\pi/2$-rotation on the target-module communication-qubit that takes $\ket{g} \rightarrow \ket{g} + \ket{e}$ and $\ket{e} \rightarrow \ket{g} - \ket{e}$.
		\begin{equation}
		\begin{aligned}
		\ket{\psi}_{\CNOT} 
		&= a \left( + \ket{01gg} - \ket{01ge} + \ket{00eg} + \ket{00ee} \right)\\
		&+ b \left( + \ket{00gg} - \ket{00ge} + \ket{01eg} + \ket{01ee} \right)\\
		&+ c \left( + \ket{10gg} + \ket{10ge} + \ket{11eg} - \ket{11ee} \right)\\
		&+ d \left( + \ket{11gg} + \ket{11ge} + \ket{10eg} - \ket{10ee} \right)
		\end{aligned}
		\end{equation}
	\end{enumerate}
	
	Now we measure the communication qubits, and we write the outcomes $\ket{g}\rightarrow ``0"$ and $\ket{e} \rightarrow ``1"$.
	\begin{equation}
	\begin{aligned}
	``00": &\ket{\psi_{00}} = a \ket{01} + b \ket{00} + c \ket{10} + d \ket{11} \\
	``01": &\ket{\psi_{01}} = -a \ket{01} + b \ket{00} + c \ket{10} + d \ket{11} \\
	``10": &\ket{\psi_{10}} = a \ket{00} + b \ket{01} + c \ket{11} + d \ket{10} \\
	``11": &\ket{\psi_{11}} = a \ket{00} + b \ket{01} - c\ket{11} - d \ket{10}
	\end{aligned}
	\end{equation}
	
	And the feedforward single-qubit operations bring the states to
	\begin{equation}
	\begin{aligned}
	``00":\xspace &\hat{IX} \ket{\psi_{00}} 
		\begin{aligned}
				  &= a \ket{00} + b \ket{01} + c \ket{11} + d \ket{10} \\
				  &= \hat{U}_\text{\CNOT} \ket{\psi} 
	     \end{aligned}\\
	``01":\xspace &\hat{ZX} \ket{\psi_{01}} = \hat{U}_\text{\CNOT} \ket{\psi} \\
	``10":\xspace &\hat{II} \ket{\psi_{10}} = \hat{U}_\text{\CNOT} \ket{\psi} \\
	``11":\xspace &\hat{ZI} \ket{\psi_{11}} = \hat{U}_\text{\CNOT} \ket{\psi}
	\end{aligned}
	\end{equation}
	
	And each of these implements a \CNOT operation.
	
	\section{Experimental toolbox} \label{sec:toolbox}
	
	\subsection{Data qubit encodings} \label{sec:data_encodings}
	
	\paragraph{Binomial encoding} 
	As discussed in the Main text, we demonstrate the teleported gate using one of the Binomial quantum codes \cite{Michael2016}, with basis states:
	\begin{equation}
	\ket{0}_L = \ket{2},\qquad \ket{1}_L = \frac{1}{\sqrt{2}}\left(\ket{0} + \ket{4}\right).
	\end{equation}
	This logical encoding provides the ability to perform quantum error correction against single photon loss events (e.g. the application of $\hat{a}$), which is the dominant error mechanism for a cavity functioning as a quantum memory. 
	A photon loss event on a quantum state $\ket{\psi_L} = \alpha \ket{0_L} + \beta \ket{1_L}$ transforms the state to $\ket{\psi_E} = \hat{a} \ket{\psi_L} = \alpha \ket{E_0} + \beta \ket{E_1}$, with error codewords $\ket{E_0}= \ket{1}$ and $\ket{E_1} = \ket{3}$. Crucially, the quantum amplitudes $\alpha$ and $\beta$ are left unchanged despite the loss event.
	Extraction of this single-photon loss error syndrome is straightforward as this error results in a photon-number parity flip from even to odd, which is readily measured in our cQED system using photon-number parity measurements \cite{Vlastakis2013}.
	Upon detection of an error event, in principle, a correction unitary can be applied that takes $\ket{E_0} \rightarrow \ket{0_L}$ and $\ket{E_1} \rightarrow \ket{1_L}$, preserving the relative quantum amplitudes $\alpha$ and $\beta$.
	In contrast to the cat code \cite{Leghtas2013,Mirrahimi2014,Ofek2016}, which has a cyclic error process, the binomial encoding requires a correction pulse after each detection of a photon jump.
	However, the binomial code provides exactly orthogonal basis states at a low photon number, thus satisfying approximate quantum error correction conditions with a lower average photon number (e.g. $\bar{n} = 2$) compared to the cat code, which requires $\bar{n} \approx 2.3$ \cite{Michael2016}. 
	In addition, the encoding into a superposition of a finite number of Fock states simplifies the control of binomial codes as compared to the infinite superposition of Fock states that define the cat code.
	The discussion over these trade-offs highlights the utility of a hardware-efficient approach, and allows for selection of logical encodings dependent on the particular application.
	
	\paragraph{Fock encoding}
	We also demonstrate the teleported gate using a simple Fock encoding, with basis states $\ket{0}$ and $\ket{1}$, utilizing the lowest two energy levels of the cavity to specify the data qubit. We note that this basis is not a logical-encoding according to our definition as it does not allow for quantum error correction; however, by specifying the data qubits in this simple basis, we can extract an upper-bound to the performance of the teleported gate using our current device.
	
	\subsection{Local operations} \label{sec:local_ops}
	
	\subsubsection{Implementation by optimal control}
	The teleported \CNOT gate requires quantum control over the data qubit (cavity) and communication qubit (transmon) within each module. We utilize a numerical technique described in detail in Ref.~\cite{Heeres2017} to design universal operations between the two qubits in each module. In particular, we use the Gradient Ascent Pulse Engineering (GRAPE) algorithm to find a set of time-dependent pulses that implements a particular unitary operation or set of quantum state transfers. Our goal is to use GRAPE to find an operation that acts on a subspace of the complete Hilbert space.
	We supply the algorithm with a set of $K$ initial and target states $\lbrace \ket{\psi_k^{init}} \rightarrow \ket{\phi_k^{target}} \rbrace$ for a given drift Hamiltonian $H_0$ and set of $M$ control Hamiltonians $\lbrace H_m \rbrace$. The GRAPE algorithm determines a set of $M$ pulses $\lbrace \epsilon_m(t) \rbrace$ with length $T$ to define an operation which maximizes the fidelity over the set of state transfers:
	\begin{equation}
	F_{OC} = \frac{1}{K^2} \left| \sum_k^K \braket{\phi^{target}_k}{\hat{U}_{OC} \psi_k^{init}} \right|^2,
	\end{equation}
	where the calculated operation, $\hat{U}_{OC}$, for a given set of control pulses is
	\begin{equation} \label{eq:U_OC}
	\hat{U}_{OC}(\lbrace \epsilon_m \rbrace) = \int_0^T \exp\left[-i (H_0 + \sum_m^M \epsilon_m(t') H_m) / \hbar \right] dt'.
	\end{equation}
	
	For each module, we implement optimal control pulses by specifying complex-valued driving terms on the cavity (data) and transmon (communication) qubits: $\epsilon_c(t) \hat{c}^\dagger +\epsilon^*_c(t) \hat{c}$ and $\epsilon_q(t) \hat{q}^\dagger + \epsilon^*_q(t) \hat{q} \approx \epsilon_q(t) \left(\op{g}{e} + \op{e}{g} \right)$, where we have taken a two-level approximation for the transmon qubit. In our numerical optimization we discretize the pulses in 2 ns time steps, taking $\epsilon_m(t) \rightarrow \epsilon_m(t_i)$. In order to accurately reflect the dynamics of these complex pulses, it is important to use many levels of the cavity mode; here, we generate all pulses using a Hilbert space of $23$ levels for the cavity and $2$ levels for the transmon.
	
	Practically, there are physical limitations to the pulses set by the control hardware; to take these effects into consideration we include three constraints to the optimization routine.
	First, we apply an amplitude penalty to ensure that the pulse drive amplitude never exceeds a threshold value.
	Second, we apply a derivative penalty that gives preference to smooth pulses and lower bandwidth pulses. 
	Finally, we include a final contraint to ensure that the pulse starts and ends with an near-zero amplitude to eliminate the possibility of sharp transients generated by pulse edges.
	

	\subsubsection{Defining local operations}
	
	We can group the pulses required for this work into three categories: single-qubit operations on the encoded data (cavity) qubit, entangling data-communication (cavity-transmon) operations, and cavity-transmon encoding/decoding pulses. For clarity we use the subscripts $c$ and $q$ to denote the cavity and transmon state, respectively.
	\begin{enumerate}
		\item The single-qubit operations, $\hat{U}^{s}$, are specified with the following state transfers: $\ket{0_L}_c \ket{g}_q \rightarrow \hat{U}^{s} \ket{0_L}_c \ket{g}_q$ and $\ket{1_L}_c \ket{g}_q \rightarrow \hat{U}^{s} \ket{1_L}_c \ket{g}_q$.
		Here, we only specify that the transmon qubit begins and ends in the ground state, an assumption we make in our experiment to allow easier generation of pulses.
		
		\item The teleported \CNOT operation requires two different entangling operations.
		First, the entangling local operation for the module containing the control data qubit (D1) requires a cavity-controlled, transmon-target local \CNOT, requiring the state transfers
		\begin{equation} \nonumber
		\begin{aligned}
		\ket{0_L}_c \left( \alpha \ket{g}_q + \beta \ket{e}_q \right) &\rightarrow \ket{0_L}_c \left( \alpha \ket{g}_q + \beta \ket{e}_q \right) \\
		\ket{1_L}_c \left( \alpha \ket{g}_q + \beta \ket{e}_q \right) &\rightarrow \ket{1_L}_c \left( \beta \ket{g}_q + \alpha \ket{e}_q \right). 
		\end{aligned} 
		\end{equation}
		In practice, this operation closely matches a photon-number parity operation, requiring only transmon pulses in a Ramsey-like operation that flips the quantum amplitudes $\alpha$ and $\beta$ for the transmon qubit if the cavity is in $\ket{1_L}$. Our numerically-defined does reflect this intuition, though generally the extracted pulses will be quite complex.
		
		Second, the entangling local operation for the module containing the target data qubit (D2) requires a transmon-controlled, cavity-target local \CNOT, requiring the state transfers
		\begin{equation} \nonumber
		\begin{aligned}
		\left( \alpha \ket{0_L}_c + \beta \ket{1_L}_c \right) \ket{g}_q &\rightarrow \left( \alpha \ket{0_L}_c + \beta \ket{1_L}_c \right) \ket{g}_q \\
		\left( \alpha \ket{0_L}_c + \beta \ket{1_L}_c \right) \ket{e}_q &\rightarrow \left( \beta \ket{0_L}_c + \alpha \ket{1_L}_c \right) \ket{e}_q.
		\end{aligned} 
		\end{equation}
		Here, the logical cavity state is flipped ($\ket{0_L}_c \leftrightarrow \ket{1_L}_c$) when the transmon qubit is in the excited state $\ket{e}_q$, a more challenging operation due to the nontrival transitions among different Fock states.
		
		To aid the GRAPE algorithm in finding an acceptable solution, we expand the search to include those that perform the desired unitary up to single-qubit $\hat{Z}$-phase freedoms on both the qubit and the cavity. 
		This generalization can offer, in some cases, dramatic speed-ups in computation time, while only requiring simple modifications in the pulse sequence.
		
		\item The start of every experiment requires preparation of the cavities into a given initial state. We implement an encoding pulse takes an arbitrary transmon state and maps it onto the encoded cavity state:
		\begin{equation} \nonumber
		\left( \alpha \ket{g}_q + \beta \ket{e}_q \right)  \ket{0}_c \rightarrow \ket{g}_q \left( \alpha \ket{0_L}_c + \beta \ket{1_L}_c \right).
		\end{equation}
		Thus, to prepare a general logical cavity state, we intialize the ground state in both the transmon and cavity, perform a simple single-qubit transmon rotation for the desired initial state, and apply the encoding pulse to load the state onto the cavity, also returning the transmon to the ground state.
		
		In order to perform logical tomography on the cavity states, we apply decoding pulses that map the logical cavity state onto the transmon qubit, essentially reversing the decoding pulse resulting in the following state transfer
		\begin{equation} \nonumber
		\left( \alpha \ket{0_L}_c + \beta \ket{1_L}_c \right) \ket{g}_q \rightarrow \ket{0}_c \left( \alpha \ket{g}_q + \beta \ket{e}_q \right).
		\end{equation}
	\end{enumerate}
	
	The simulated fidelities for all of our optimal control operations are in excess of $\mathcal{F} > 99\%$. Note that this optimization does not include the presence of loss and decoherence as we numerically integrate the Schrodinger equation. After developing these pulses, we apply them in a full Lindblad master equation simulation that accounts for these nonidealities to extract an expected operation fidelity; typically these fidelities are measured to be $1-3\%$ lower, depending on the pulse length. We provide an exhaustive list in \autoref{table:oct_fidelities}.
	
	\subsubsection{Experimental calibration}
	
\begin{figure*}[tp]
		\centering{
			\phantomsubcaption\label{subfig:oct_rb_circuit}
			\phantomsubcaption\label{subfig:oct_rb_data}
			\phantomsubcaption\label{subfig:oct_qpt_circuit}
			\phantomsubcaption\label{subfig:oct_qpt_data}
		}
		\includegraphics[width=120mm]{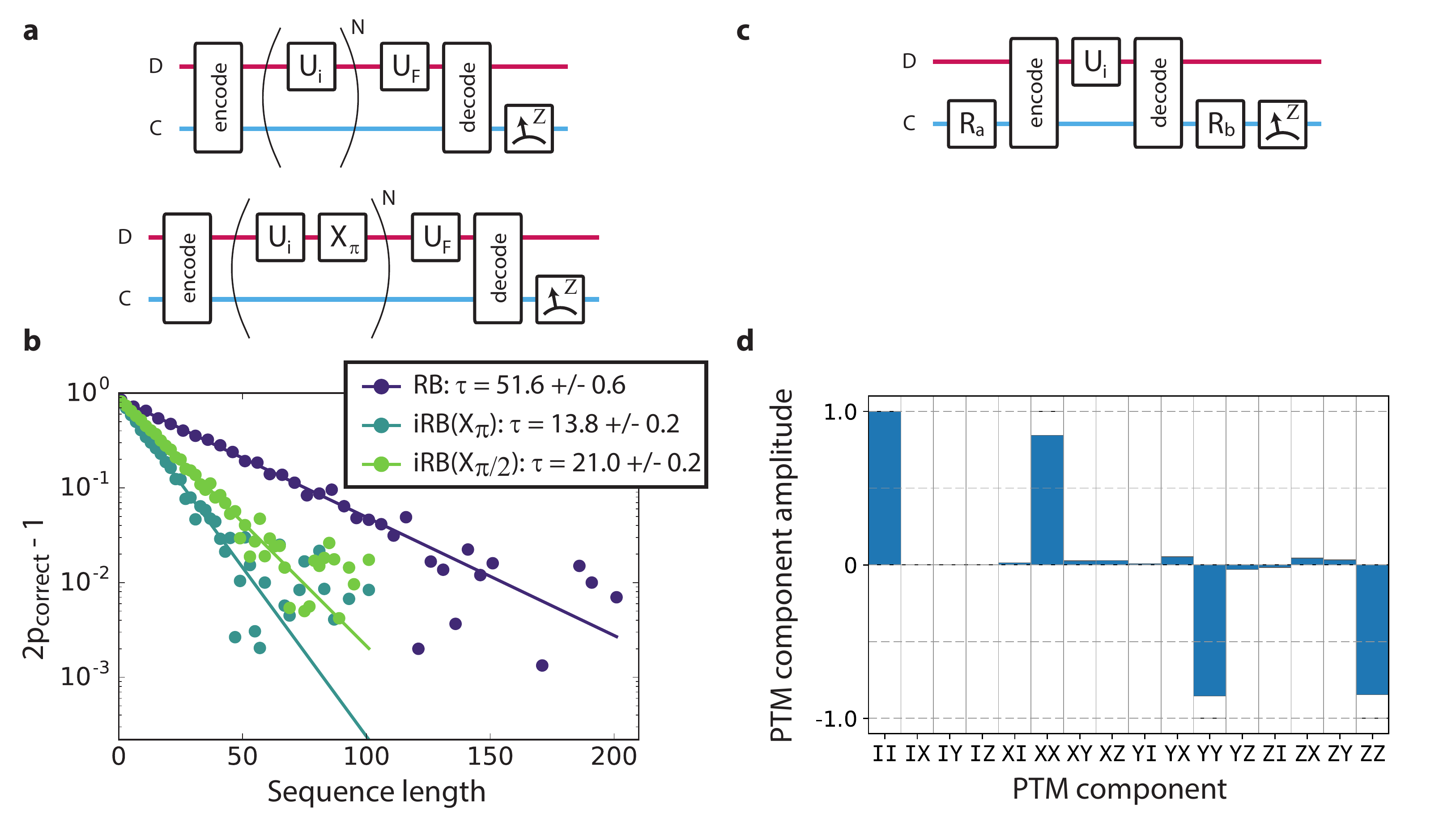}
		\caption{ \label{fig:oct_rb_qpt}
			\textbf{Characterization experiments for optimal control pulses.}
			\subref{subfig:oct_rb_circuit}
			Pulse sequences for randomized benchmarking (RB, top) and iterative randomized benchmarking (iRB, bottom).
			For RB, a sequence of $N$ operations ($\hat{U}_i$) are randomly applied to the qubit state and a final correction unitary ($\hat{U}_F$) ideally inverts the effect of the composite sequence.
			Measurement of the resulting state is then compared to the expected state to establish the error per operation.
			Here, to utilize RB for characterization of logical operations, the standard RB protocol is modified with an encode pulse before and a decode pulse after the RB sequence.
			For iRB, to characterize a particular operation (here, $U_X$), this operation is interleaved among the random operations.
			A comparison with the standard RB sequence allows extraction of the single operation fidelity.
			In our implementation of RB, a new random sequence is generated for every experimental realization (or shot) and for each length $N$, $\hat{U}_F$ is chosen to ideally flip the state to both $\ket{0}_L$ and $\ket{1}_L$.
			\subref{subfig:oct_rb_data}
			Typical results for RB and iRB experiments.
			We plot data (dots) for a scaled probability of measuring the correct result as a function of the number of random pulses, $N$, in the RB sequence.
			These data are fit to the following model: $p_\text{correct} = 0.5 + A e^{-\tau / N}$. 
			From this fit, we estimate an error per gate to be: $r = (1 - e^{-1/ \tau(RB)}) / 2$.
			From these fits, we extract an average gate error for an $\hat{X}$ operation: $r(X) = (1 - e^{-1/ \tau(X) - 1/\tau{RB}}) / 2$.
			\subref{subfig:oct_qpt_circuit}
			Characterization of optimal control operation $U_i$ using process tomography.
			To characterize this logical operation we perform communication qubit QPT for the operation $\hat{U}_\text{encode} \hat{U}_i \hat{U}_\text{encode}$.
			For communication qubit QPT, we perform a set of transmon rotations before ($\hat{R}_a$) and after ($\hat{R}_b$) the operation under test.
			\subref{subfig:oct_qpt_data}
			Typical results for QPT experiments, here an $X_\pi$ logical gate.
			We present the results in the Pauli transfer representation, with each bar $AB$ (with $A,B \in \lbrace \hat{I}, \hat{X}, \hat{Y}, \hat{Z}\rbrace$) representing each element in the Pauli transfer matrix (PTM). The experimentally reconstructed process is shown in blue; the ideal process is shown in hollow, black outlined bars.
		}
	\end{figure*}

	Successful implementation of these optimal control pulses in experiment relies on two broad requirements: first, an accurate knowledge of the drift Hamiltonian, $H_0$, which has been discussed in \autoref{sec:hamiltonian}; and second, a careful characterization of the experiment control lines. Here, we focus on the second requirement and detail our tune-up protocol for these optimal control pulses. We calibrate five parameters for each module's optimal control pulses: drive amplitudes for the cavity and transmon drives, linear frequency-dependent amplitude dispersion for both drives, and a relative timing between the two drives. Notionally, a single set of tuning parameters maximizes the fidelity for all optimal control pulses in a given module; therefore, for simplicity, we choose to optimize single-qubit gates. The method used here closely matches the approach in Ref. \cite{Heeres2017}; we perform randomized benchmarking (RB) to extract an metric related to average gate fidelity to optimize over the set of five parameters. 
	Having performed calibrations we proceed to extract experimental optimal control pulse gate fidelities. We utilize interleaved randomized benchmarking (iRB) and quantum process tomography (QPT) to establish the performance of our operations (\autoref{fig:oct_rb_qpt}). 
	From these experiments, we extract single-qubit gate fidelities between $96-98\%$ and provide a comparison of measured and simulated gate fidelities are given in \autoref{table:oct_fidelities}.
	In general, our experimental results are consistent with the simulated fidelities and demonstrate that this technique is a powerful tool to implement complex operations on a logical qubit encoded in the levels of a cavity.
	
	Improvements on our implementation of the teleported \CNOT will depend critically on the quality of these optimal control operations. As such, a careful analysis of the specific types of errors that may occur during the local operations is important. Our simulations indicate that a large fraction of the errors result in codespace leakage (e.g. the final state is no longer in the logical subspace of the data qubit with the communication qubit in the ground state, $\left( \alpha \ket{0_L}_c + \beta \ket{1_L}_c \right) \ket{g}_q$). The optimal control operation takes the joint state through a complex trajectory in its Hilbert space; an error will induce a new trajectory that will affect the final state, likely resulting in population outside of the logical subspace \cite{Heeres2017}. From typical simulations we roughly quantify that ${\sim}95\%$ of the total infidelity is due to this leakage; the fidelity of the operation within the codespace is ${>}99\%$. Though the root case of this error may be due to a combination of factors (here the performance is limited by transmon coherence), the result can be characterized by this single error syndrome. Therefore, an outstanding question to further improve the performance of these optimal control operations will be whether leakage detection circuits can be designed to efficiently herald when these errors occur.

	\begin{table*}[htp] 
		\begin{ruledtabular}
			\begin{tabular}{c c l l l l}
				Encoding	& Module	& Operation	($\hat{U}$)	& Pulse length ($\si{\us}$) 	& Predicted infidelity ($\%$) & Inferred experimental infidelity ($\%$) \\
				&			&						&				& $(1 - \mathcal{F}_\text{sim})$ 	
				& $\Delta\mathcal{F} = (\mathcal{F}_\text{E+D} - \mathcal{F}_\text{expt})$ \\
				\hline \noalign{\vskip 1pt}
				Binomial	& 1 		& $\hat{U}_\text{E+D}$ (Encode \& decode)	& 1.2 each	& 6.9	& $(1 - \mathcal{F}_\text{E+D}) = 7.1$ \\
				& 			& $\hat{X}_{\pi}$							& 1.4		& 3.7 	& 2.4	\\
				& 			& $\hat{X}_{\pm\pi/2}, \hat{Y}_{\pm\pi/2}$	& 1.4-1.5	& 3.5	& 3.2	\\
				& 			& $\hat{U}_{\CNOT,1}$						& 0.6		& -- 	& -- 	\\
				Binomial	& 2 		& $\hat{U}_\text{E+D}$ (Encode \& decode)	& 1.2 each	& 4.4	& $(1 - \mathcal{F}_\text{E+D}) = 5.3$ 	\\
				& 			& $\hat{X}_{\pi}$							& 1.0		& 2.4 	& 2.1	\\
				& 			& $\hat{X}_{\pm\pi/2}, \hat{Y}_{\pm\pi/2}$	& 1.0 - 1.5	& 2.3 	& 2.6 	\\
				& 			& $\hat{U}_{\CNOT,2}$						& 2.0		& 5.4	& -- 	\\
				\hline\noalign{\vskip 1pt}
				Fock		& 1 		& $\hat{U}_\text{E+D}$ (Encode \& decode)	& 0.6 each	& 4.1	& $(1 - \mathcal{F}_\text{E+D}) = 3.1$	\\
				& 			& $\hat{X}_{\pi}$							& 0.7		& 1.6 	& 1.4	\\
				& 			& $\hat{X}_{\pm\pi/2}, \hat{Y}_{\pm\pi/2}$	& 0.7		& 1.8	& 1.3	\\
				& 			& $\hat{U}_{\CNOT,1}$						& 0.9		& -- 	& -- 	\\
				Fock		& 2 		& $\hat{U}_\text{E+D}$ (Encode \& decode)	& 0.6 each	& 2.3	& $(1 - \mathcal{F}_\text{E+D}) = 2.5$ 	\\
				& 			& $\hat{X}_{\pi}$							& 0.9		& 1.3 	& 0.7	\\
				& 			& $\hat{X}_{\pm\pi/2}, \hat{Y}_{\pm\pi/2}$	& 0.7		& 1.2 	& 0.6 	\\
				& 			& $\hat{U}_{\CNOT,2}$						& 1.0		& 2.9	& --	\\
			\end{tabular}
		\end{ruledtabular}
		\caption{\label{table:oct_fidelities}
			\textbf{Local operation infidelities.}
			Predicted and experimental results for optimal control pulses encoded in the Binomial and Fock bases.
			The predicted infidelity ($1 - \mathcal{F}_\text{sim}$) is taken from a time-domain simulation that includes finite coherences of the transmon and cavity.
			The experimental infidelity is calculated from the difference of two separate QPT experiments.
			We first perform process tomography on only the encode-decode process, $\hat{U}_\text{dec}U_\text{enc}$, to extract $\mathcal{F}_\text{E+D}$ for each module and encoding. Each experiment that involves the cavity requires an encoding and decoding pulse. We then perform QPT on the process that includes the operation under test, $\hat{U}$, and extract a process fidelity $\mathcal{F}_\text{expt}$ on the combined process $\hat{U}_\text{dec} \hat{U} \hat{U}_\text{enc}$. Note that this fidelity includes the contribution of the encode and decode operations. We report the difference of the two experiments to estimate the fidelity of the target operation alone: $\Delta\mathcal{F} = (\mathcal{F}_\text{E+D} - \mathcal{F}_\text{expt})$.	
		}
	\end{table*}

	\subsection{Generation of communication qubit Bell pair} \label{sec:comm_bell_pair}
\begin{figure*}[tp]
		\centering{
			\phantomsubcaption\label{subfig:rip_gate_circuit}
			\phantomsubcaption\label{subfig:rip_gate_expt}
			\phantomsubcaption\label{subfig:rip_gate_sweep}
			\phantomsubcaption\label{subfig:bell_gen_circuit}
			\phantomsubcaption\label{subfig:qubit_bell_qst}
		}
		\includegraphics{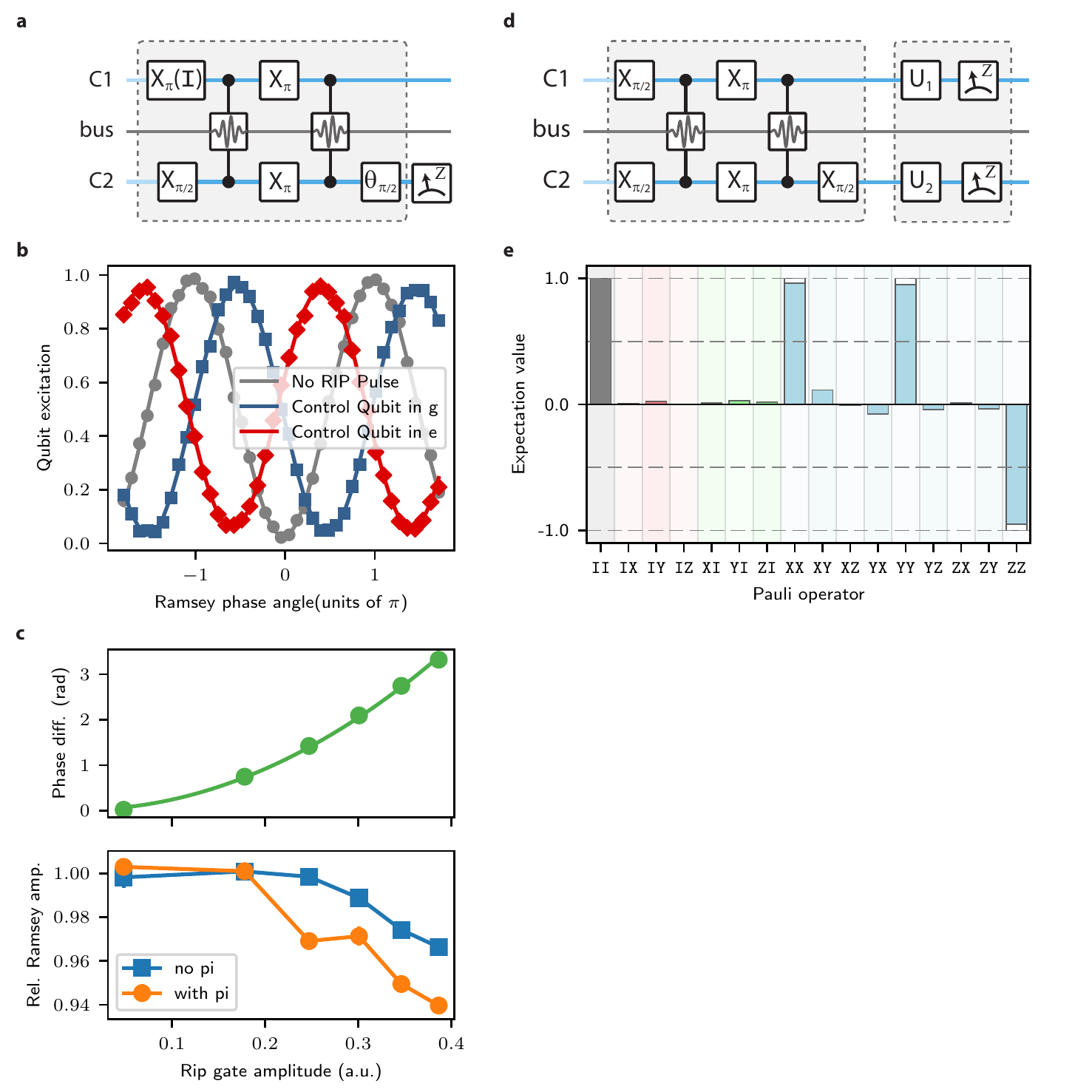}
		\caption{ \label{fig:rip_gates_bell_qst}
			\textbf{Generation of communication qubit Bell pair.}
			\subref{subfig:bell_gen_circuit} Pulse sequence for tuning up RIP gates.
			A refocused RIP sequence is used to entangle the two communication qubits, explicitly including the bus. This experiment to calibrate the RIP pulse amplitude, detuning, and length are similar to a Ramsey phase-style experiment. To measure the entangling phase, the target qubit (C1) is initialized either in $\ket{e}$ or $\ket{g}$ by performing a $\pi$-pulse or identity, the RIP sequence is performed with varying angle ($\theta_{\pi/2}$) on C2 for the second $\pi/2$-pulse, and finally, C2 is measured.
			\subref{subfig:rip_gate_expt} Typical refocused-RIP Ramsey phase experiment.
			Characteristic sinusoidal oscillations of the target qubit state are observed while varying the angle $\theta_{\pi/2}$. The phase of the oscillations depend on the communication qubit initialization (blue square for $\ket{g}$ and red diamonds for $\ket{e}$), and the phase difference between the two experiments is a hallmark of an entangling phase. Here, the RIP gate amplitude is set to achieve a $\pi$ phase difference between the two RIP experiments. A reference experiment where the RIP pulse amplitude is set to zero is shown (grey circles); we observe only a small reduction in amplitude when the RIP pulses are included as compared to this reference experiment. 
			\subref{subfig:rip_gate_sweep} RIP gate amplitude sweep.
			The experiment in \subref{subfig:rip_gate_expt} is performed for several RIP gate amplitudes and the extracted entangling phase is extracted (top) as well as the relative oscillation amplitude as compared to the reference (bottom). The entangling phase is ideally proportional to the RIP gate power ($\phi \propto \epsilon^2$), and a quadratic fit is shown indicating that the data is consistent with the expected trend.
			\subref{subfig:bell_gen_circuit} Pulse sequence for generating the communication qubit Bell pair.
			After generating the Bell state (first block), QST is performed on both of the qubits to assess the quality of the entangled state.
			\subref{subfig:qubit_bell_qst} Characterizing communication qubit Bell pair.
			Experimentally measured Pauli vector components of the two communication qubits. The generated state is $\left(\ket{ge}+ \ket{eg}\right) / \sqrt{2}$ with the ideal denoted as hollow bars. For this reconstruction, we take $10,000$ averages per tomography setting.
		}
	\end{figure*}

	The teleported \CNOT protocol begins with the entanglement of the two communication qubits; in this section, we provide additional details on our implementation of this sequence.
	As discussed in the Main Text, we use a resonator induced phase (RIP) gate to generate a Bell pair between the two communication qubits \cite{Cross2015, Paik2016}.
	We apply an off-resonant, shaped drive $\epsilon(t)$ with a carrier frequency detuned from the bus resonance frequency by $\Delta_0 \approx \SI{20}{MHz}$ which induces a qubit-state dependent Stark shift on the bus.
	This drive induces a displacement of the bus oscillator: $\xi_{\Delta} = -\frac{i \epsilon_d}{2 \left( i\Delta + \kappa / 2 \right)}$. 
	The size of this displacement is dependent on the detuning between the drive frequency and the qubit-state dependent cavity frequency; within the two-transmon manifold $\ket{ij}$, where $i,j \in \lbrace g, e \rbrace$, the corresponding detunings given as $\Delta_{ij} = \Delta_0 + \chi_{ij}$, where $\chi_{ij}$ corresponds to the total dispersive shift for state $\ket{ij}$. 
	In our experiment, we operate in the low-loss regime where $\kappa \ll \Delta$ so that $\xi_{\Delta} \approx -\frac{\epsilon_d}{2\Delta}$.
	This bus displacement gives rise to a Stark shift of the transmon frequency by an amount that is dependent nonlinearly with the total dispersive shift, $\delta_{ij} = \chi_{ij} |\xi_{ij}|^2$.
	This Stark shift generates a transmon-state dependent phase: $\phi_{ij} = \int \delta_{ij}(t) dt$.
	The quantity that describes the magnitude of entanglement is
	$\phi_{ent} = \phi_{ee} - \phi_{eg} - \phi_{ge} - \phi_{gg}$, which extracts the non-separable two-qubit phase contribution from the single qubit phase contributions.
	For approximately equal $\chi$, we can simplify this expression $\phi_{ent} = \frac{\chi^2}{2\Delta^3} \int \left|\epsilon_d(t) \right|^2 dt$.
	We detail our tuneup protocol in \autoref{fig:rip_gates_bell_qst}.
	
	In our experiment, we utilize a RIP pulse of length $T = \SI{300}{ns}$ of the form: $\epsilon(t) = A \left[ \cos \left(\pi \cos \left(\pi t/T\right) \right) + 1 \right]$ in order to minimize residual photon population left in the bus cavity at the end of the pulse \cite{Paik2016}.
	We implement a refocused-RIP sequence that includes two RIP pulses that sandwich a $\pi$-pulse; this sequence is utilized to ``echo'' away the always-on dispersive interaction between the communication-qubit (transmon) and its data-qubit (cavity).
	We are able to achieve an entangling phase of $\phi_{ent} = \pi$ in $672$ ns, and combined with single communication qubit rotations, create the Bell state $\ket{\Phi^+} = \left( \ket{ge} + \ket{eg} \right) / \sqrt{2}$ with state fidelity of $\left( 97\pm1\right)\%$. 
	Our error bar is a average of several experiments and roughly accounts for systematic errors in our experiment; statistical errors are smaller in this experiment, ${<}1\%$, as extracted from a bootstrap analysis.
	In \autoref{subfig:bell_gen_circuit} and \autoref{subfig:qubit_bell_qst}, we show the experimentally measured two qubit Bell state.
	
	\subsection{Communication qubit measurement and reset} \label{sec:comm_meas_reset}
	The success of the teleported \CNOT requires reliable measurements of each communication qubit. As discussed previously our JPC-enabled single-qubit readout has assignment fidelities in excess of $99\%$. In our implementation of the teleported gate, the communication qubits serve dual roles: both to store inter-module entanglement and also to enable complex data qubit operations via optimal control pulses. Therefore, after the measurement of the communication qubits in our protocol, we perform a feedback reset of both communication qubits to the ground state to recycle them for the following single-qubit operations and tomography steps. These measurements are required to be highly-quantum non-demolition to both the communication qubit as well as the data qubits.
	
\begin{figure*}[tp]
		\centering{
			\phantomsubcaption\label{subfig:transmon_reset_circuit}
			\phantomsubcaption\label{subfig:transmon_reset_qst_cond}
			\phantomsubcaption\label{subfig:transmon_reset_qst_all}
		}
		\includegraphics{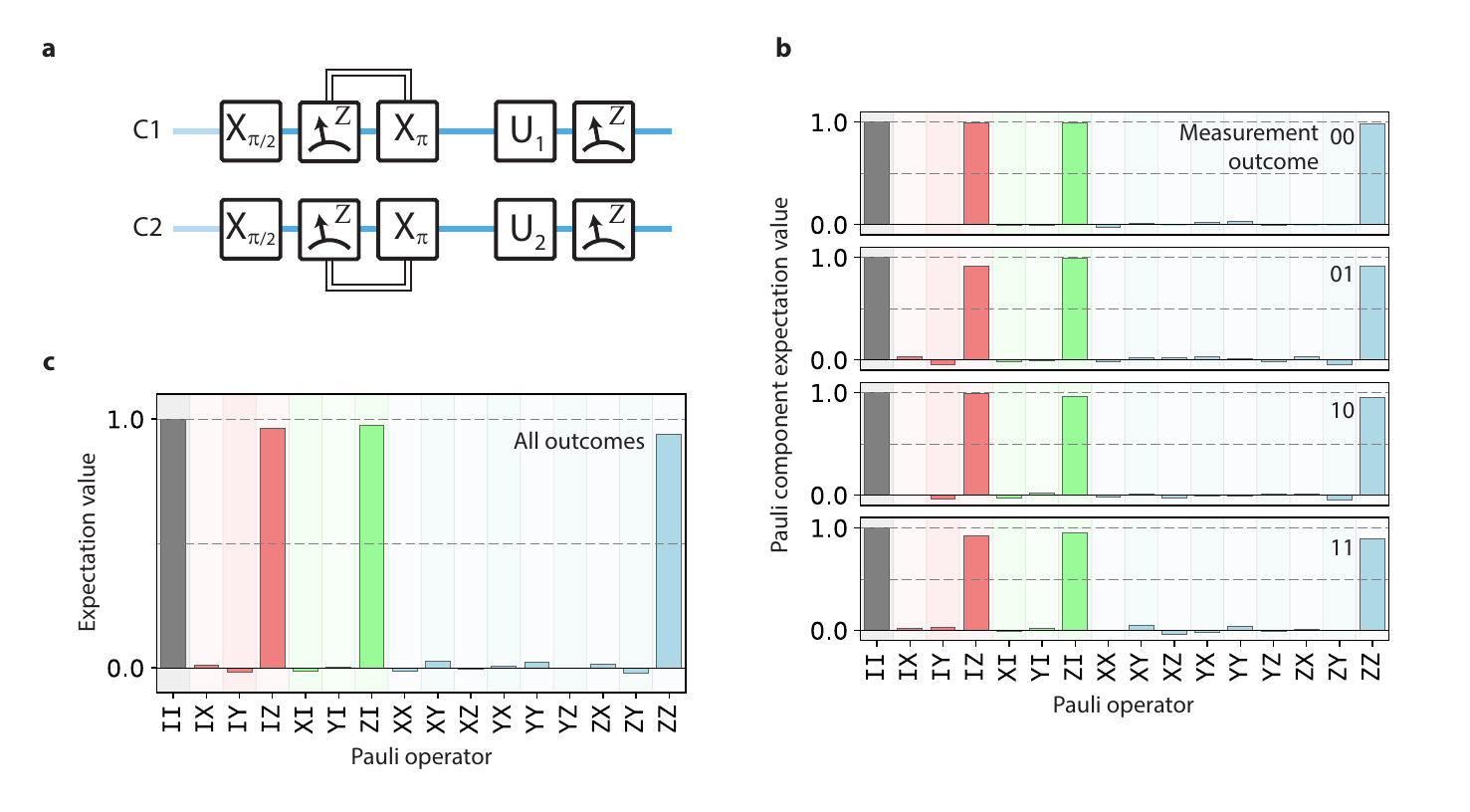}
		\caption{ \label{fig:transmon_reset}
			\textbf{Communication qubit measurement and reset.}
			\subref{subfig:transmon_reset_circuit} Pulse sequence for testing communication qubit measurement and reset.
			The two communication qubits (transmons) are initialized in the joint state $\left(\ket{gg} + \ket{ge}+ \ket{eg} + \ket{ee}\right)/2$. The two qubits are then measured and if the measurement indicates that the state is projected to $\ket{e}$ a $\pi$-pulse is applied to flip the state to the ground state. Conditional QST is performed to analyze the quality of measurement and reset. This measurement and reset protocol is used in the teleported gate.
			\subref{subfig:transmon_reset_qst_cond} Experimentally measured Pauli vector components conditioned on the measurement outcome. 
			We assign a ``0'' (``1'') to indicate that the measurement projected the qubit to be in $\ket{g}$ ($\ket{e}$). For all outcomes, we find high fidelity to the two-qubit ground state, $\ket{gg}$ as expected with ground state fidelities $\lbrace 00: 99.3\%, 01: 95.7\%, 10: 97.7\%, 11: 94.2\% \rbrace$. From these results, we establish that the measurement and feedback processes for each qubit are independent; from the single-qubit reset infidelities, we expect a measurement fidelity of $1 - \left(0.993 - 0.957 \right) - \left( 0.993 - 0.957 \right) = 0.948$, which is consistent with the result for measurement outcome 11.
			\subref{subfig:transmon_reset_qst_all} Experimentally measured state after measurement-based reset.
			Measurement results from \subref{subfig:transmon_reset_qst_cond} are combined, and the compiled results illustrate that the reset protocol is high-fidelity and independent of measurement outcome. The fidelity of this reconstructed two-qubit state to $\ket{gg}$ is $96.9\%$. 
		}
	\end{figure*}

	We perform the following experiment to test both the measurement as well as the reset. First, we initialize the two communication qubits in an equal superposition of computational states: $\ket{\psi_{init}} = \left(\ket{gg} + \ket{ge} + \ket{eg} + \ket{ee} \right) / 4$. Next, we perform measurements on each qubit allowing the control computer to perform real-time state estimation. Conditioned on the measurement results, we apply a $\pi$-pulse if the qubit was measured to be in the excited state. Finally, we analyze the state via conditioned state tomography to assess the quality of the reset. The resulting tomograms are shown in \autoref{fig:transmon_reset}. 
	We extract state infidelities to the joint ground state $\ket{gg}$ of ${<}1\%$ for the case when we measured both qubits in the ground state, outcome ``00''. 
	We observe single-qubit infidelities of $2\%$ and $4\%$ when each qubit is measured to be in the excited state.
	The result from outcome ``11'' indicates that these infidelities are additive and any crosstalk in the measurement or control is negligible.
	From these results, we find an average reset infidelity of ${\sim}3\%$, primarily limited by decay during the measurement and subsequent controller latency.
	From this experiment we establish that our system enjoys highly accurate and QND single-qubit measurements.
	
	\section{Tuning up the teleported CNOT} \label{sec:tuneup}
	
	\subsection{Logical vs. Reference phases}
	
	When manipulating logical qubits, it is necessary to distinguish two types of phase shifts: logical phase shifts and reference-frame phase shifts.
	A logical phase shift is a phase shift between the two logical basis states, and is generated by the logical $\hat{Z}$ operator, $\hat{Z}_L(\phi) = \text{diag}\left[1, e^{i\phi}\right]$:
	\begin{equation}
	\begin{aligned}
	\alpha \ket{0_L} + \beta \ket{1_L} &\longrightarrow \alpha \ket{0_L} + \beta e^{i\phi} \ket{1_L} \\
	&= \alpha \ket{2} + \frac{\beta}{\sqrt{2}} e^{i\phi} \left( \ket{0} + \ket{4} \right),
	\end{aligned}
	\end{equation}
	
	On the other hand, a reference-frame phase shift is generated by the phase-shift operator, $\hat{U}_\text{ref.}(\theta) = \exp\left[ -i \theta \hat{a}^\dagger \hat{a} \right]$, which acts on the physical levels of the state,
	\begin{equation}
	\begin{aligned}
	\alpha \ket{0_L} + \beta \ket{1_L} &\longrightarrow \hat{U}_\text{ref.}(\theta) \left[ \alpha \ket{0_L} + \beta \ket{1_L} \right] \\
	&= e^{i 2 \theta}\left[ \alpha \ket{2} + \frac{\beta}{\sqrt{2}} \left(e^{-i2\theta} \ket{0} + e^{i2\theta} \ket{4} \right) \right]
	\end{aligned}
	\end{equation}
	
	While $\hat{Z}$-phases and reference phases are equivalent for a single physical qubit (and this equivalence has been previously utilized to implement $\hat{Z}$-gates through software reference phase updates \cite{McKay2017}), these phases have distinct effects on a logical qubit state.
	In particular, for our cQED system, the dispersive interaction, $H_I = \chi \hat{a}^\dagger \hat{a} \op{e}$, naturally generates a reference-frame phase shift when the communication-qubit is in the $\ket{e}$ state; for a time $t$, the total phase accumulation is given as $\theta = \chi t$.
	Correct determination of these reference phases shifts are critical to the successful application of our optimal control pulses; these pulses must be applied with the correct phase relative to the logical Bloch sphere for the data-qubit.
	In the following sections, we detail our tune-up protocol, establishing how we keep track of the reference phases necessary for the implementation of the teleported \CNOT gate.
	Crucially these phases are determined either from direct measurements or from the Hamiltonian and pulse sequence timing, and are known in advance.
	
	\subsection{Reference phases due to Bell state generation} \label{sec:tuneup_local_ops}
	
	We consider the reference-phase shift induced by the Bell state generation to determine the phase-adjustment necessary for the local \CNOT operations.
	This is necessary as the data-qubit states encode quantum information prior to generation of the communication-qubit Bell pair as would be typical in any algorithm that uses the teleported \CNOT. 
	During the Bell generation step, each communication-qubit induces a reference frame shift on its respective data-qubit according to their dispersive interaction.
	Our Bell generation is similar to a spin-echo sequence, and so the length of time that the communication qubit is in the excited state half time for half of the duration of the operation, or $T_\text{Bell} / 2 = \SI{336}{ns}$. This results in estimated phase shifts of the control and target data-qubit of $\SI{1.21}{rad}$ and $\SI{1.78}{rad}$, respectively.
	
\begin{figure*}[tp]
		\centering{
			\phantomsubcaption\label{subfig:bell_phase_circuit}
			\phantomsubcaption\label{subfig:bell_phase_data}
		}
		\includegraphics[width=100mm]{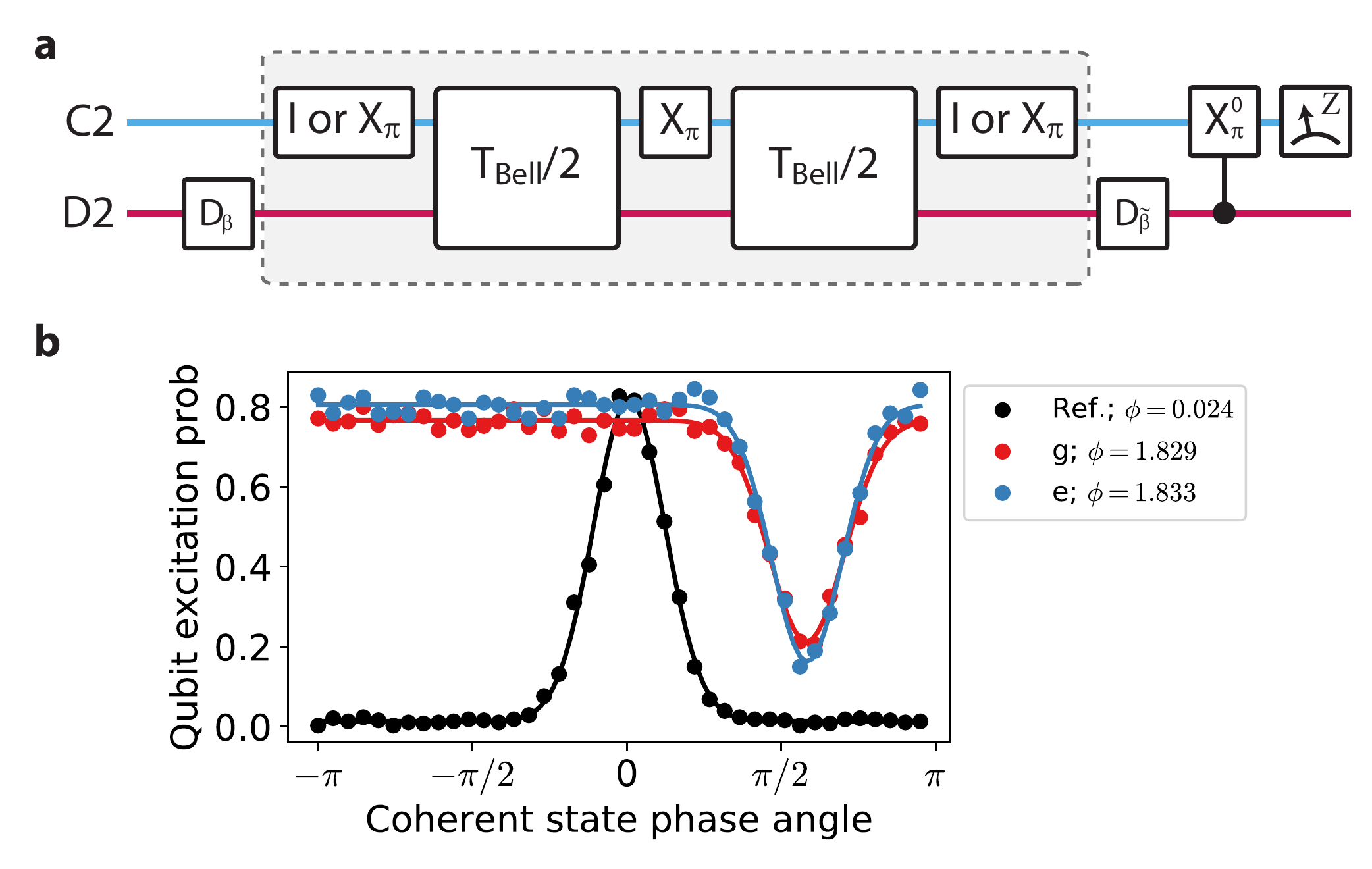}
		\caption{ \label{fig:bell_ref_phase}
			\textbf{Measurement reference phase due to Bell state generation.}
			\subref{subfig:bell_phase_circuit}
			Pulse sequence for measuring Bell state generation reference phase shift.
			Pulses $D_\beta$ indicate data qubit displacements of $\beta$; $X_\pi$ indicate communication qubit $\hat{X}$-rotations of angle $\theta$; $X_\pi^0$ indicates a communication qubit rotation of angle $\pi$ that is selective on the data qubit having zero photons. The block labeled $T_\text{Bell}$ indicates a time delay during which the data qubit undergoes a communication qubit state-dependent phase shift. In this sequence we either initialize the communication qubit in $\ket{g}$ or $\ket{e}$ by performing identity (I) or a $\pi$-pulse ($\hat{X}_\pi$). 
			The shaded section indicates the sequence that is similar to the Bell state generation protocol. For this experiment, the second displacement is related to the initial displacement by $\tilde{\beta} = -\beta e^{i\phi}$.
			\subref{subfig:bell_phase_data}
			Experimental results indicating the final phase of the data qubit coherent state as a function of phase angle.
			Data are presented as dots; gaussian fits to the data are shown as lines.
			The center of the gaussian peak (or dip) represents the phase of the coherent state.
			We perform a reference experiment (Ref, in black) that removes the shaded section of the pulse sequence, and we measure a peak at phase angle $0$, as expected.
			Results where we initalize the communication qubit state in $\ket{g}$ (g, in red) or $\ket{e}$  (e, in blue) illustrate an average reference phase shift of $\phi = \SI{1.831}{rad}$.
			Importantly, extracted phases for both experiments closely match, indicating that the final state of the data qubit does not depend on initial state of the communication qubit.
		}
	\end{figure*}

	For a more realistic measurement of this phase shift, we perform the following experiment that takes into consideration finite pulse timings and other experimental details (\autoref{fig:bell_ref_phase}).
	Here, we utilize a pulse sequence similar to the Bell generation protocol; in this case, we remove the bus drives to keep the communication qubits in a separable state.
	We first displaced the data qubit (cavity) to the coherent state $\ket{\alpha} = 2$.
	Then, we perform a sequence similar to the Bell state generation, except we use a bus drive amplitude of zero to keep the two communication qubits separable.
	Next, we either initialize the communication qubit in $\ket{g}$ or $\ket{e}$ as two separate experimental variations. 
	Then, we continue with the sequence, and at the end of the protocol, perform a $\pi$ pulse if the communication qubit was initialized in $\ket{e}$. 
	In this way, for both results, the communication qubit will be left in the excited state (the ground state could also be used).
	These two experiments allow us to check that the refocusing $\pi$-pulse does indeed disentangle the data qubits by the end of the sequence.
	At the end of the protocol, the data qubit state will have evolved from $\ket{\alpha} \rightarrow \ket{\alpha e^{i\phi}}$ where $\phi = \chi T_\text{Bell} / 2$, and we measure $\phi$ by performing a series of data qubit displacements $D_\alpha = 2$ with varying phases. 
	We extract phase shifts of $\phi = 1.81$ for both communication qubit initial states, indicating that the data and communication qubit will be left separable at the end of the Bell state generation protocol.
	We find that our experimentally phase is in close agreement ($\sim1\%$) with the simple estimation.
	
	\subsection{Reference phases due to measurements} \label{sec:tuneup_meas}
	Next, we consider the references phases accumulated during the communication-qubit measurements. These phases, in contrast to the previous section, are now conditioned on the measurement outcome. When the communication-qubit is measured in $\ket{g}$, the data-qubit acquires no additional reference phase; however, when the communication-qubit is measured in $\ket{e}$, then the data-qubit acquires a total reference phase $\theta_{M} = \chi T_M$, 
	where $T_M$ and is the duration of the total measurement process, including measurement pulse, integration, and state estimation.
	
\begin{figure*}[tp]
		\centering{
			\phantomsubcaption\label{subfig:ref_phase_msmts_phase}
			\phantomsubcaption\label{subfig:ref_phase_msmts_wigner}
		}
		\includegraphics{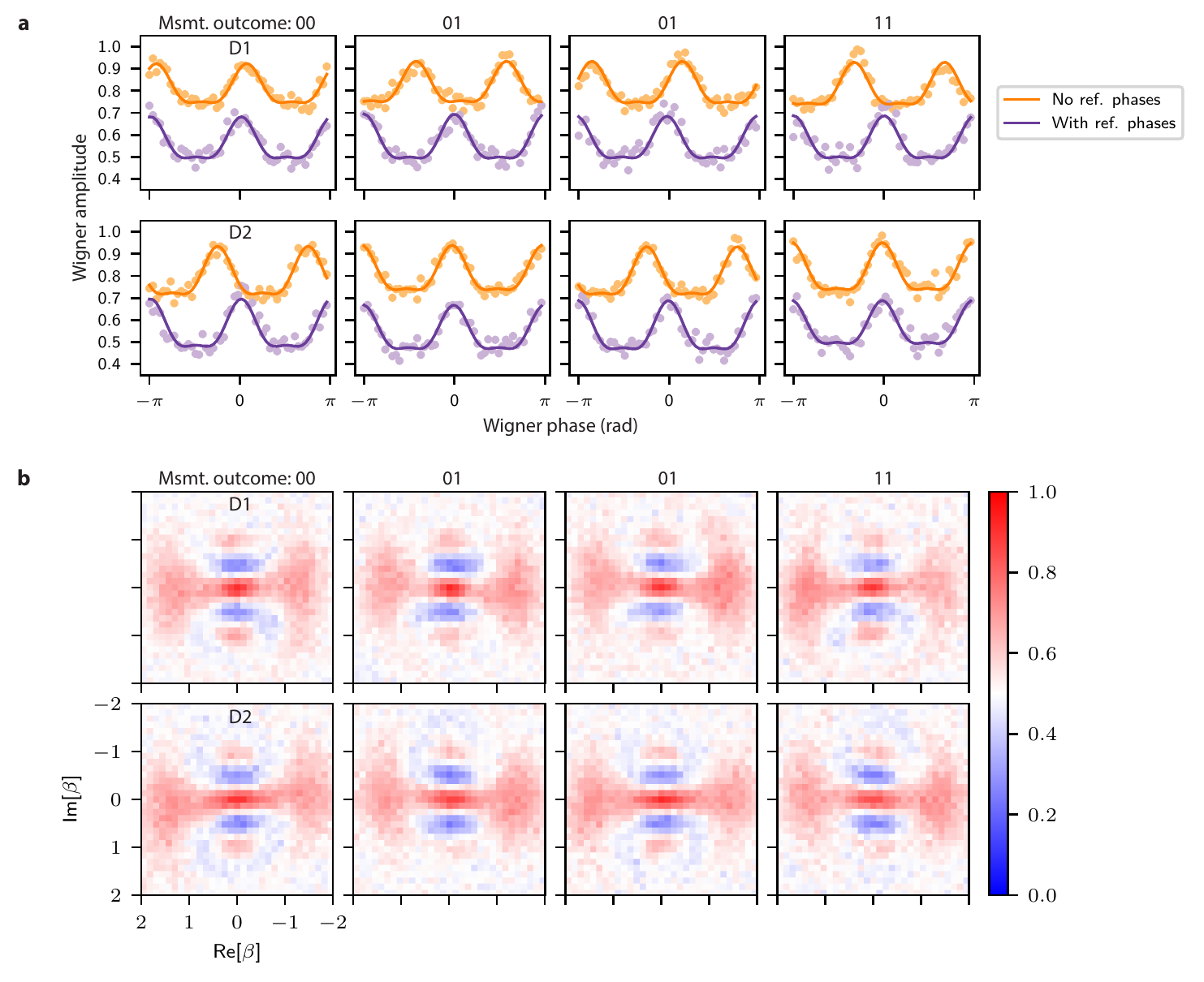}
		\caption{ \label{fig:tuneup_meas_phase}
			\textbf{Measurement-induced reference phase shift.}
			For all data, we perform the teleported \CNOT on the maximal superposition state $\left(\ket{0_L} + \ket{1_L}\right)_{D1} \otimes \left(\ket{0_L} + \ket{1_L}\right)_{D2} / 2$. The \CNOT is invariant on this particular input state and so the output state is ideally separable, which allows analysis of the output state via single qubit Wigner tomography. 
			The ideal state for each data qubit should be $\ket{{+}X_L} = \left(\ket{0_L} + \ket{1_L}\right) / \sqrt{2}$, a horizontally-oriented kitten state.
			For both \subref{subfig:ref_phase_msmts_phase} and \subref{subfig:ref_phase_msmts_wigner}, the top row represents the result from data qubit 1 (the control qubit) and the bottom row represents the result from data qubit 2 (the target qubit).
			Each column represents a different measurement outcome of the communication qubits.
			\subref{subfig:ref_phase_msmts_phase}  One-dimensional phase-cut of the Wigner function for the output state. In each panel, the top data (orange) represents case when the reference phases are not accounted for, and we observe different reference phase shifts for each measurement outcome. The solid line represents a fit to the data for target $\ket{{+}X_L}$ rotated in the IQ space. Taking these reference phases into consideration, applying them in real-time, we find the bottom data (purple), where we observe that all measurement outcomes have the same reference phase. The data without accounting for reference phases (orange) is vertically offset for clarity.
			\subref{subfig:ref_phase_msmts_wigner} Wigner function of the output state when accounting for reference phases. For all outcomes, the target data qubit state, $\ket{{+}X_L}$, is generated with the appropriate reference phase.
		}
	\end{figure*}

	To experimentally extract the measurement-outcome dependent phases, we prepared the input state $\ket{\psi_{in}} = \ket{{+}\hat{X}_L}\ket{{+}\hat{X}_L}$, and applied the teleported \CNOT. 
	Here, the \CNOT is invariant to the input state; thus, the output state should remain as the separable state $\ket{\psi_{in}}$.
	We extract the resulting state and perform Wigner tomography on each data-qubit.
	The effect of these reference phase shifts will induce a rotation in the IQ-plane of the state and can be parameterized by $\theta$: $\ket{{+}\hat{X}_L(\theta)} = \left(\ket{0} + e^{i2\theta} \sqrt{2}\ket{2} + e^{i4\theta} \ket{4}\right) / \sqrt{2}$.
	The resulting Wigner functions (\autoref{fig:tuneup_meas_phase}) are then used to extract a set of eight phases (one phase for each measurement outcome, for each data-qubit) to account for the reference frame shift induced by the measurement. 
	
	Due to the probabilistic nature of the communication measurements, our controller performs a critical task to store these phases in memory, selecting the correct phase depending on the measurement outcome for each experimental shot.
	In \autoref{fig:tuneup_meas_phase}, we account for this reference frame shift and note the extracted Wigner functions are all correctly aligned.
	The extracted phases should be considered as a persistent reference frame update that is applied to all subsequent operations on the data qubits; here, we apply these phases to both the feedforward operations as well as the decoding operations.
	In our encoding, a logical $\hat{Z}_L$ phase-flip where $\phi = \pi$ is equivalent to a reference phase shift of $\theta = \pi / 2$. Therefore, we apply the feedforward $\hat{Z}$ operation in software by conditionally updating the phase of the cavity drive for the control module. 
	
	
	\subsection{Communication qubit measurement basis} \label{sec:tuneup_meas_basis}
	
\begin{figure*}[tp]
		\centering{
			\phantomsubcaption\label{subfig:tuneup_meas_op_circuit}
			\phantomsubcaption\label{subfig:tuneup_meas_op_data}
		}
		\includegraphics[width=130mm]{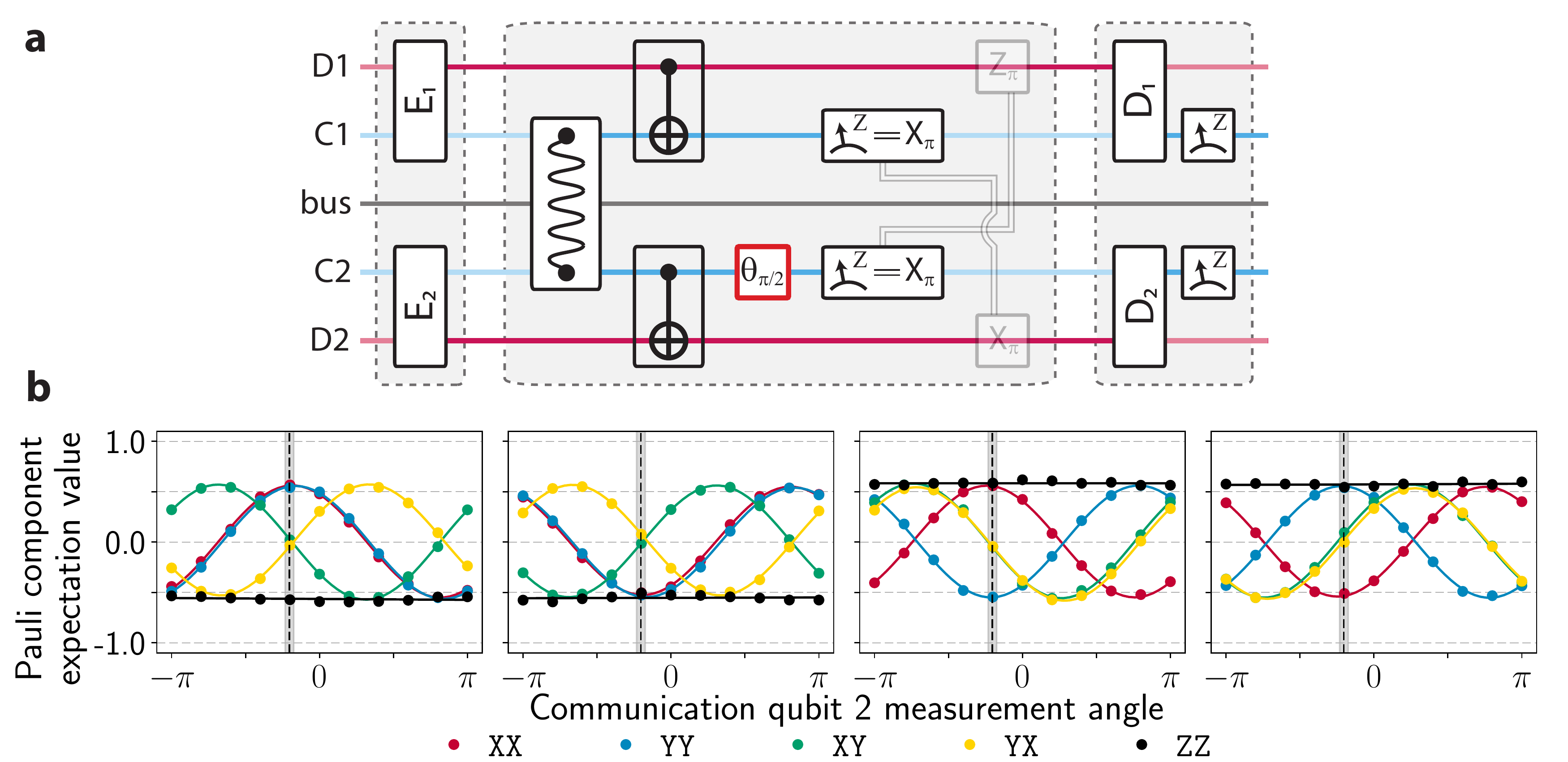}
		\caption{ \label{fig:tuneup_meas_op}
			\textbf{Communication qubit 2 measurement angle.}
			\subref{subfig:tuneup_meas_op_circuit} Pulse sequence for tuning up C2 measurement angle. 
			The experiment is similar to \autoref{fig:teleported_cnot}, with two main modifications. First, in this experiment, the $\pi/2$-pulse on C2 (outlined in red) is varied; second, the feedforward operations are left out. For each measurement angle, we perform conditioned QST on the output state.
			\subref{subfig:tuneup_meas_op_data} Measured Pauli components as a function of C2 measurement angle, conditioned on measurement outcome. The experiment described in \subref{subfig:tuneup_meas_op_data} is performed on the input state $\left(\ket{0_L} + \ket{1_L}\right)\ket{0_L}$ and the teleported operation is performed while varying angle $\theta$. Results from density matrix reconstruction is shown as dots, and a fit to the data is shown as lines. The optimal measurement angle to implement the teleported \CNOT operation is highlighted with the dashed vertical line with uncertainty in its value as the shaded grey region.
		}
	\end{figure*}

	In contrast to the reference phase shifts induced by the dispersive interaction as discussed in the previous two sections, the choice of basis of the communication qubit measurements can induce a logical phase and therefore plays an important role in determining the exact operation of the teleported gate. In particular, we study the effect of changing the measurement basis on C2, which notionally should be a $\hat{X}$ measurement. In the following experiment (\autoref{fig:tuneup_meas_op}), we run the teleported gate while sweeping the phase of the $\pi/2$-pulse on C2 prior to the communicaton qubit measurements (outlined in red in \autoref{subfig:tuneup_meas_op_circuit}). The C2 measurement operator is given generally as $\hat{M}(\phi) = \cos(\theta) \hat{X} + \sin(\theta) \hat{Y}$, where $\theta$ is the chosen angle of the $\pi/2$ pulse; that is, we are rotating the measurement basis around the equator of the Bloch sphere of C2. Ideally, $\theta = 0$ to achieve the desired $\hat{X}$ measurement; however, we expect an offset in this measurement angle due to single qubit phases acquired during the previous local operation (where the optimal control pulse induces reference phase shifts on both the data and communication qubit). 
	
	In our experiment, we perform the teleported gate on the input state, $\ket{\psi_\text{in}} = \left(\ket{0_L} + \ket{1_L}\right)\ket{0_L}$ while varying the measurement angle $\theta$. For simplicity, we perform the experiment without feedforward operations and extract conditioned QST for each measurement angle. For the measurement angle that corresponds to an $\hat{X}$-measurement, we expect to generate the following conditioned states: $\lbrace \ket{\Psi^+_L}, \ket{\Psi^-_L}, \ket{\Phi^+_L}, \ket{\Phi^-_L}\rbrace$ (if we had added feedforward operations, then we would ideally generate the even Bell state $\ket{\Phi^+_L}$). Our results are illustrated in \autoref{subfig:tuneup_meas_op_data}, where we plot selected Pauli operators. For each measurement outcome, we find that the two-qubit parity $\expval{ZZ}$ is conserved over all measurement angles indicating that we generate a maximally entangled state independent of angle; however, we observe oscillations in the transversal two qubit operators, $\expval{XX}, \expval{XY}, \expval{YX}, \expval{YY}$. These oscillations are expected and indicate that the choice of measurement angle induces a logical phase on the output state. We also observe that the contrast of the output state is constant over all measurement angles. From this experiment, we extract the optimal measurement angle to implement the teleported \CNOT operation. The observations from this experiment suggests that the choice of measurement basis may also allow for tuning of the particular teleported operation.
	
	\section{Teleported \CNOT experimental data} \label{sec:cnot_results}
	
	In this section, we expand upon the teleported \CNOT process results as discussed in the Main Text. 
	In \autoref{sec:exp_gate_fidelity}, we describe additional details on the \CNOT fidelity.
	In \autoref{sec:error_budget}, we describe the error budget that is described in the Main Text.
	
	\subsection{Teleported \CNOT performance} \label{sec:exp_gate_fidelity}
	
	\autoref{fig:extended_qpt_binomial} and \autoref{fig:extended_qpt_fock} show comprehensive process results for the teleported \CNOT gate for both binomial and Fock encodings, respectively. In additional to the compiled results, we also provide the measurement outcome-conditioned processes to further highlight the role of classical communication and feedforward for the teleported gate.
	
	From these results, we quantify the performance of the teleported gate by calculating the process fidelity of the entire operation. We describe two quantities: 
	\begin{enumerate}
		\item $\mathcal{F}_\text{pro}$: The process fidelity is extracted from the combined operation $\mathcal{E}_\text{all} = \hat{U}_\text{dec} \hat{U}_\text{\CNOT}\hat{U}_\text{enc}$, and includes the effect of the encoding and decoding pulses in the fidelity.
		\item $\mathcal{F}_\text{gate}$: The inferred gate fidelity is calculated from $\mathcal{F}_\text{pro}$ relative to the fidelity of only the encode and decode pulses, $\mathcal{F}_\text{E+D}$. Explicitly, $\mathcal{F}_\text{gate} = \mathcal{F}_\text{pro} + (1 - \mathcal{F}_\text{E+D})$. 
	\end{enumerate}
	Additional details on the tomography method is provided in \autoref{sec:tomography}.
	
	As described in \autoref{sec:local_ops}, the encoding and decoding processes are accomplished through the use of optimal control pulses. 
	The fidelities of the encode and decode operations are taken from \autoref{table:oct_fidelities},
	\begin{quote}
		Encode/decode (Binomial): \\
		$1 - \mathcal{F}_\text{E+D} = (6.9 + 4.4)\% = 11.3\%$
		
		Encode/decode (Fock): \\
		$1 - \mathcal{F}_\text{E+D} = (4.1 + 2.3)\% = 6.4\%$
	\end{quote}
	Here, we use the simulated process fidelities to conservatively estimate the gate fidelity; the experimentally extracted fidelities would result in a slightly high teleported \CNOT gate fidelity.
	The compiled experimental results for the teleported \CNOT are provided in \autoref{table:fidelities_expt}. 
	
	
	%
	%
	
	\begin{table*}[tp]
		\begin{ruledtabular}
			\begin{tabular}{l c c c c c c c}
				& & \multicolumn{4}{c}{Measurement outcome, $\mathcal{F}_\text{pro}$ ($\mathcal{F}_\text{gate}$), ($\%$) } \\
				\noalign{\vskip 1pt} \cline{3-6} \noalign{\vskip 1pt}	
				Encoding	& Feedforward (FF) 	& 00			& 01			& 10	& 11	& All \\
				\hline \noalign{\vskip 1pt}
				Binomial	& w/ FF				& 68 (79)	& 67 (78)	& 70 (81)	& 68 (79)	& 68 (79) \\
				& no FF				& 71 (82)	& 68 (80)	& 70 (82)	& 68 (79)	& N/A \\
				\hline \noalign{\vskip 1pt}
				Fock		& w/ FF				& 82 (88)	& 79 (86)	& 81 (88)	& 79 (86)	& 80 (87) \\
				& no FF				& 83 (89)	& 80 (87)	& 80 (87)	& 78 (84)	& N/A \\
			\end{tabular}
		\end{ruledtabular}
		\caption{\label{table:fidelities_expt}
			\textbf{Experimental process fidelities.} 
		}
	\end{table*}
	
	\subsection{Error budget} \label{sec:error_budget}
	
	In this section we provide further details on the estimates of gate error as presented in the Main Text. 
	Our approach reflects and expands upon an error model described in Ref.~\cite{Jiang2007}: the total gate error is given as the additive error contribution from each element of the teleported gate. 
	We consider each element independently, estimating the total loss from a combination of experimentally measured and simulated quantities.
	Therefore, our estimate of the gate fidelity accounts for all known non-idealities of the system, providing an upper bound in the actual experimental performance.
	This estimate provides a useful benchmark to assess the potential for other unknown sources of loss.
	The error bars reported for each element are estimated from systematic run-to-run variation.
	
	\subsubsection{Communication qubit Bell pair}
	The communication qubit Bell pair is generated with the same pulse sequence independent of data qubit encoding. 
	Our characterization of the Bell pair in \autoref{sec:comm_bell_pair} is performed when both data qubits are not encoded and left in the vacuum state, and serves as an upper bound to the Bell pair fidelity when used in the teleported \CNOT.
	We set the error probability of the communication qubit Bell pair from this bare state fidelity, $p_\text{Bell}$:
	\begin{quote}
		Comm. qubit Bell pair:\\
		$p_\text{Bell} = 1 - \mathcal{F}_\text{Bell} = \left(3 \pm 1 \right) \%$
	\end{quote}
	When the Bell state generation sequence is applied during the teleported gate, the encoded data qubit state will induce a dispersion in the communication qubit frequency, potentially affecting the quality of communication qubit pulses. To account for this, we use short pulses ($\sigma = \SI{6}{ns}$) which have a bandwidth approaching two orders of magnitude larger than this dispersion.
	In principle, it is possible to characterize the communication qubit Bell pair when the data qubits are initialized in an encoded state; in practice, the presence of data qubit photons makes it difficult to perform reliable state tomography. As such, we use the above quantity for the purpose of this calculation.
	
	\subsubsection{Local operations}
	We perform a local operation within each module, which are implemented via optimal control pulses.
	For the control module (module 1), we perform a local \CNOT that is controlled by the data qubit, targeting the communication qubit. 
	This operation is a photon-number parity mapping for the Fock encoding \cite{Vlastakis2013} and a ``super-parity'' (e.g. $0 \mod 4$ vs. $2 \mod 4$) mapping for the Binomial encoding; as such, the operation time for the binomial encoding ($\SI{500}{ns}$) is about half the length as compared to the Fock encoding ($\SI{900}{ns}$). The operation fidelities for the binomial and Fock encodings are $2\%$ and $3\%$, respectively (\autoref{table:oct_fidelities}).
	For the target module (module 2), the \CNOT is now controlled by the communication qubit, targeting the data qubit. This is a highly nontrivial operation: when the communication qubit is in $\ket{e}$, the operation flips the logical basis states, $(\ket{0} + \ket{4}) / \sqrt{2} \leftrightarrow \ket{2}$. As a result, in order to achieve a high fidelity operation, we use a $\SI{2000}{ns}$ optimal control pulse.
	The total local operation infidelity for the binomial and Fock encodings are $\left( 5 \pm 3\right) \%$ and $\left(3 \pm 2\right)$, respectively (\autoref{table:oct_fidelities}). The errors are estimated from the range of measured coherence times.
	
	In addition to the infidelity associated with each optimal control pulse, it is also important to consider the timing of the two pulses (\autoref{subfig:telecnot_timing}). 
	In our experiment, both pulses start at the same time, but can have different pulse lengths.
	This alignment results in a delay between the local operation and the communication qubit measurement for module 1, during which the communication qubit may suffer a $T_1$ error. 
	Importantly, $T_2$ errors are projected out during the subsequent $\hat{Z}$ measurement and do not contribute to the infidelity of this component.
	This decay will cause an error in the measurement outcome assignment in the subsequent measurement, introducing additional infidelity to the $00$ and $01$ paths while reducing the probability of measuring outcomes $10$ and $11$.
	For the binomial and Fock encodings, we estimate this additional infidelity to be $p_{T_1} \approx 0.025$ and $0.002$, respectively.
	Therefore, we can describe the path-dependent error probabilities, $p_{LO,i}$ for $i\in \lbrace 00, 01, 10, 11 \rbrace$: 
	\begin{quote}
		Local operations (Binomial):\\
		$p_\text{LO,i} = \left( \lbrace 10, 10, 7, 7 \rbrace \pm 3 \right) \%$ 
		
		Local operations (Fock): \\
		$p_\text{LO,i} = \left( \lbrace 6, 6, 6, 6 \rbrace \pm 2 \right) \%$
	\end{quote}
	
	\subsubsection{Communication qubit measurements}
	Measurement of the communication qubits can suffer from two general types of errors: assignment errors and measurement-decay errors. An assignment error is characterized by inferring the incorrect communication qubit state (e.g. assigning the state $\ket{eg}$ as 00 instead of 10). Measurement-decay errors represent the case when the communication qubit undergoes a relaxation event during the measurement process. In the latter case, the resulting qubit state as well as the measurement outcome may not reflect the state prior to measurement.
	For the teleported \CNOT, either error will lead to the application of incorrect communication qubit reset pulse as well as the incorrect feedforward operation in the teleported \CNOT. 
	Furthermore, if the communication qubit is not properly reset to the ground state, then all subsequent optimal control pulses, including feedfoward and decoding pulses, will fail.
	To holistically account for these errors, we use our results from \autoref{sec:comm_meas_reset}, where we measurement the fidelity of resetting the communication qubits to the ground state.
	From this experiment we find the following measurement-outcome dependent errors, $p_{Msmt,i}$ for $i\in{00, 01, 10, 11}$: 
	\begin{quote}
		Comm. qubit measurements:\\
		$p_\text{Msmt,i} = \left( \lbrace 1, 4, 2, 6 \rbrace \right) \%$ 
	\end{quote}
	
	\subsubsection{Feedforward operations}
	Depending on measurement outcome, the teleported \CNOT requires single-qubit feedfoward operations, a $\hat{Z}$ and $\hat{X}$ operation on module 1 and 2, respectively. 
	We implement the logical Z operation in software by updating the phase reference of the data qubit, which adds no time and has unit fidelity. Therefore, we only need to consider the infidelity when the $\hat{X}$ feedforward operation is applied to module 2. This operation has an infidelity of $3\%$ and $2\%$ for the binomial and Fock encoding, respectively, and is applied only for outcomes $00$ and $01$. 
	The additional error probabilities, $p_\text{FF,i}$ associated with the feedfoward operations are given as:
	\begin{quote}
		Feedforward operations (Binomial):\\
		$p_\text{FF,i} = \left( \lbrace 3, 3, 0, 0 \rbrace \pm 1 \right) \%$
		
		Feedforward operations (Fock):\\
		$p_\text{FF,i} = \left( \lbrace 2, 2, 0, 0 \rbrace \pm 1 \right)\%$
	\end{quote}
	
	\subsubsection{Total infidelity}
	From the previous sections, we extract the total gate error, $p_\text{CNOT}$, as
	\begin{equation}
	p_\text{CNOT} = p_\text{Bell} + p_\text{LO} + p_\text{Msmt} + p_\text{FF}
	\end{equation}
	The results from this section are summarized in \autoref{table:error_budget}.
	We extract conditioned errors for the teleported gate of $p_\text{CNOT, i} = \left( \lbrace 16, 20, 13, 16 \rbrace \pm 3\right)\%$ for the Binomial encoding and $p_\text{CNOT, i} = \left( \lbrace 12, 15, 12, 15 \rbrace \pm 2\right)$ for the Fock encoding.
	From these conditioned results, we expect total gate errors of $p_\text{CNOT} = \left( 16 \pm 3 \right)\%$ and $p_\text{CNOT} = \left( 13 \pm 2\right)\%$ for the binomial and Fock encoding, respectively.
	From \autoref{table:fidelities_expt}, our inferred gate infidelities are $1 - \mathcal{F}^\text{inf}_\CNOT = \left( 21 \pm  2\right)\%$ and $\left( 13 \pm 2\right)\%$, both of which are consistent with our error model.

	\begin{table*}[tp]
		\begin{ruledtabular}
			\begin{tabular}{l l c c c c c}
				& & \multicolumn{4}{c}{Measurement outcomes} \\
				\cline{3-6} \noalign{\vskip 1pt}
				Encoding	& Gate component	& 00	& 01	& 10	& 11 & All \\
				\hline \noalign{\vskip 1pt}
				Binomial	& Bell generation, $p_\text{Bell} (\%)$ 					& 3		& 3 	& 3 	& 3		& 3 \\
				& Local operations, $p_\text{LO,i} (\%)$					& 10	& 10 	& 7		& 7		& 8 \\
				& Communication qubit measurements, $p_\text{Msmt,i} (\%)$	& 1		& 4		& 2 	& 6		& 3 \\
				& Feedforward operations, $p_\text{FF,i} (\%)$				& 3		& 3		& 0.0 	& 0.0	& 2 \\
				\noalign{\vskip 2pt} \cline{3-7} \noalign{\vskip 2pt}
				& Total infidelity, $p_\text{CNOT,i} (\%)$								& 17	& 20	& 12	& 16	& 16 \\
				\noalign{\vskip 2pt} \hline \noalign{\vskip 1pt}
				Fock	& Bell generation, $p_\text{Bell} (\%)$ 						& 3		& 3 	& 3 	& 3		& 3 \\
				& Local operations, $p_\text{LO,i} (\%)$						& 6		& 6 	& 6 	& 6		& 6 \\
				& Communication qubit measurements, $p_\text{Msmt,i} (\%)$		& 1		& 4 	& 2 	& 6		& 3 \\
				& Feedforward operations, $p_\text{FF,i} (\%)$					& 2		& 2		& 0 	& 0		& 1 \\
				\noalign{\vskip 2pt} \cline{3-7} \noalign{\vskip 2pt}
				& Total infidelity, $p_\text{CNOT,i} (\%)$						& 12	& 15	& 11	& 15	& 13 \\
				\noalign{\vskip 2pt}
			\end{tabular}
		\end{ruledtabular}
		\caption{\label{table:error_budget}
			\textbf{Error budget, theory.} 
		}
	\end{table*}


	\begin{figure*}[tp]
		\centering{
			\phantomsubcaption\label{subfig:extended_qpt_binomial_noff}
			\phantomsubcaption\label{subfig:extended_qpt_binomial_wff}
		}
		\includegraphics{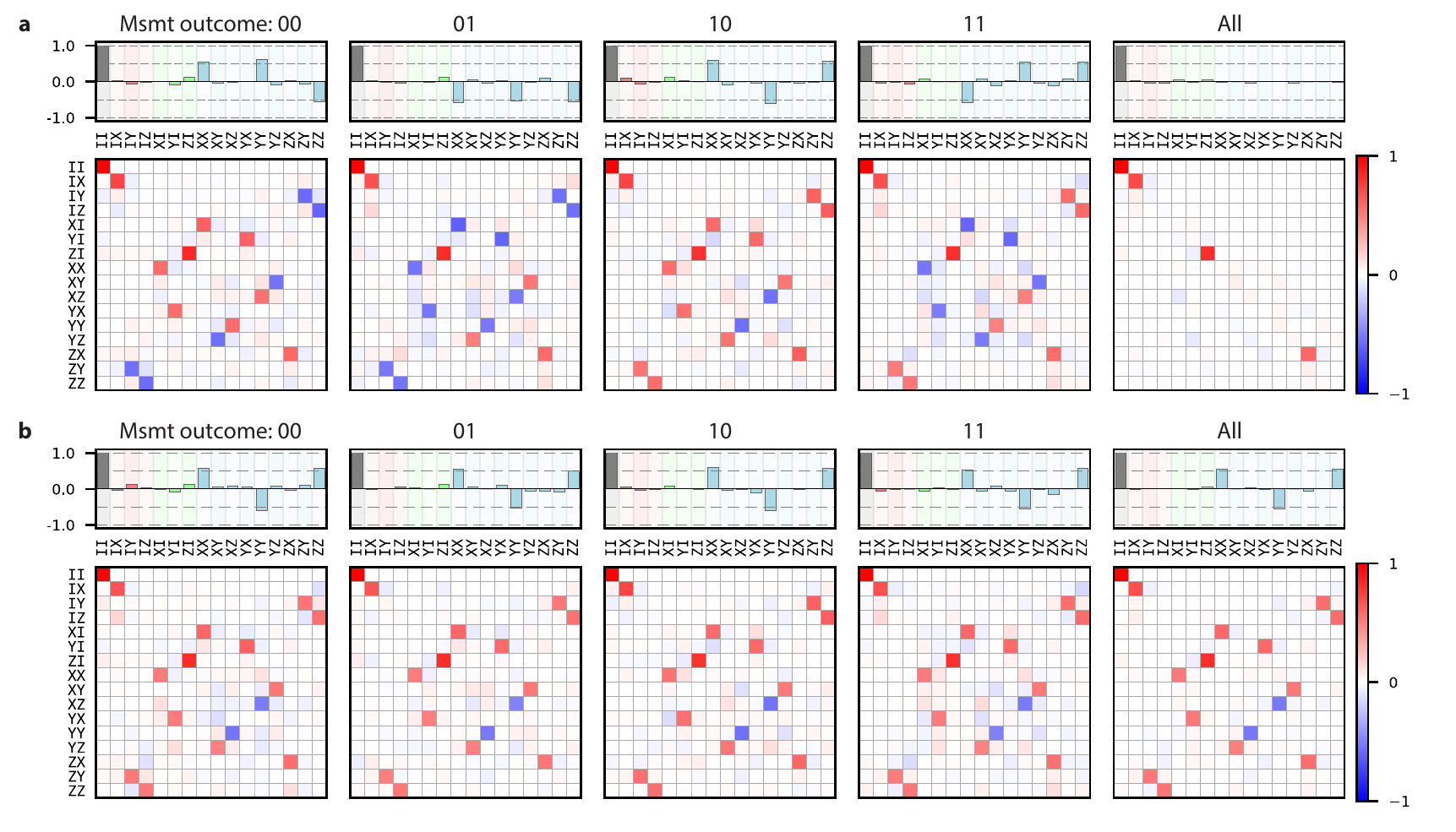}
		\caption{ \label{fig:extended_qpt_binomial}
			\textbf{Extended Binomial QPT data.}
			For each panel, we plot both the process matrix in the Pauli transfer representation (below) as well as a reconstructed state represented in the Pauli basis (above). For the reconstructed state, we choose the input state $\left(\ket{0} + \ket{1}\right)\ket{0} / \sqrt{2}$, which should result in the following Bell state when the \CNOT is applied: $\ket{\Phi^+} = \left(\ket{00} + \ket{11}\right) / \sqrt{2}$. The ideal process for each panel is represented by the dominant components taken to $\pm 1$ and small components taken to $0$.
			\subref{subfig:extended_qpt_binomial_noff} Conditioned QPT results when the feedforward operations are not applied. 	
			The first four panels (labeled: 00, 01, 10, 11) represent the process conditioned on measurement outcome. Each qualitatively has the same features (e.g. the same non-zero elements of the process matrix); however, the differing signs between the four outcome indicates that each process is modified by single-qubit operations. When all measurement results are combined (labeled: All), most of the features are washed away and only certain Pauli operators are left invariant by the process: $\lbrace II, IX, ZI, ZX \rbrace$. Notably, these operators are exactly the feedforward operations that would normally be applied. This behavior can also be observed in the state tomography results (above), where each measurement outcome heralds a different Bell state ($\lbrace \ket{\Psi^+}, \ket{\Psi^-}, \ket{\Phi^+}, \ket{\Phi^-} \rbrace$); when taken all together, the states add incoherently, resulting in a completely mixed state.
			\subref{subfig:extended_qpt_binomial_wff} Conditioned QPT results when the feedforward operations are applied. Here, all measurement outcomes (00, 01, 10, 11) indicate the same process, that of the $\CNOT$ process. Therefore, when the measurement outcomes are all taken together (All), the compiled process is that of a \CNOT gate. 
			Each tomography setting in this dataset consists of $2500$ averages; we perform a total of six pre- and post- rotations for QPT, leading to a total of $6^4 = 1296$ experiments for QPT.
		}
	\end{figure*}
	
	\begin{figure*}[tp]
		\centering{
			\phantomsubcaption\label{subfig:extended_qpt_fock_noff}
			\phantomsubcaption\label{subfig:extended_qpt_fock_wff}
		}
		\includegraphics{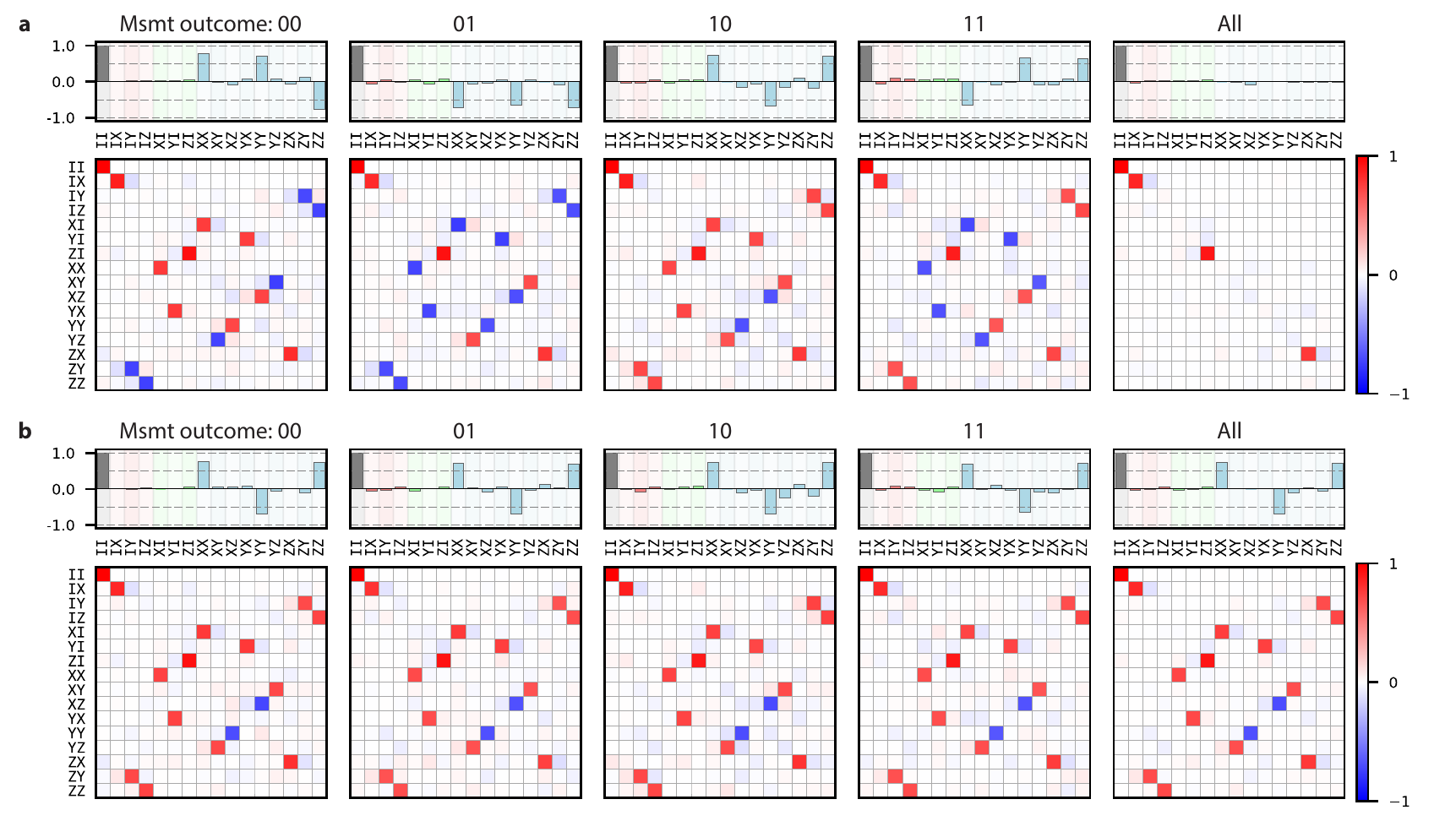}
		\caption{ \label{fig:extended_qpt_fock}
			\textbf{Extended Fock QPT data.} Data is presented in the same format as in \autoref{fig:extended_qpt_binomial}.
		}
	\end{figure*}
	
	\subsection{Contributions to infidelity}
	
	By using numerical simulations, we have estimated the relative contributions of loss that limit the performance of the teleported \CNOT. Finite $T_2$ and $T_1$ of the transmon qubits are the two dominant sources of infidelity in our experiment, accounting for roughly $70\%$ and $25\%$ of the total infidelity, respectively. In our implementation, the finite cavity lifetime accounts for $4\%$ of the total infidelity, limiting the gate fidelity at $98\%$. However, single photon errors are, in principle, detectable through implementing a quantum error correction scheme, which has not been done in this implementation. At the level of our measurement precision, we are not able to observe significant limitations due to unexpected interactions or control errors.
	
	
	\section{Analysis methods}
	
	\subsection{Quantum state and process tomography} \label{sec:tomography}
	
	Here, we discuss our methods for perform quantum state and process tomography on our system. In this work, we have performed three variants of the standard tomography protocol: 
	First, we perform transmon state tomography on the two communication qubits. 
	Second, we perform logical state tomography on the two data qubits. We use an approach described in \cite{Heeres2017} that will be further described below in \autoref{sec:logical_qst}.
	Third, we perform Wigner tomography on the cavity state to extract the full density operator of each cavity state. Despite differences in the experimental protocol, the reconstruction technique remains the same for each of the three tomography protocols.
	
	\subsubsection{Writing the tomography problem}
	In quantum state tomography, an unknown quantum state $\RhoOp$ is characterized using a set of measurements that form a complete basis of properties to reconstruct the state. This process can be written as: $\pi_k = \Tr \left[ \hat{M}_k \cdot \RhoOp \right]$, which describes measurement probability $\pi_k$ when applying the measurement operator $\hat{M}_k$ to the state $\RhoOp$. 
	In practice, however, we use a limited set of measurements operators (or POVM elements) $\lbrace \hat{E}_m \rbrace$, and instead, assemble a tomographically-complete set of operations $\lbrace \hat{U}_r \rbrace$ to act on the unknown state:
	\begin{equation} \label{eq:tomo_eqn_general1}
	\pi_{m,r} = \Tr\left[ \hat{E}_m \hat{U}_r \RhoOp \hat{U}_r^\dagger \right],
	\end{equation}
	where the set of measurements conserve probability, satisfying the completeness relation $\sum_m \hat{E}_m = \hat{I}$.
	
	To proceed, we decompose the density operator in a particular basis, generically as $\RhoOp = \sum_a^{N_a} p_a \cdot \RhoOp_a = \vec{P} \cdot \vec{\RhoOp}_a$, and applying to \autoref{eq:tomo_eqn_general1} yields
	\begin{equation} \label{eq:tomo_eqn_general2}
	\pi_{m,r} = \sum_a p_a \cdot \Tr\left[ \hat{E}_m \hat{U}_r \RhoOp_a \hat{U}_r^\dagger \right] 
	\longrightarrow
	\vec{\Pi} = \vb{T} \cdot \vec{P},
	\end{equation}
	where $\vec{\Pi}$ measurement-outcome column-vector, and $\vec{P}$ is the vector representation of the density operator, and $\vb{T}$ is the linear tomography matrix that relates the quantum state to measurement outcomes.
	Then, in principle, the quantum state can be simply reconstructed by $\vec{P} = T^{-1} \vec{\Pi}$.
	Note that for $N_m$ measurement settings and $N_r$ tomography operations, the tomography matrix has dimension $\left[N_m \cdot N_r, N_a\right]$, and is in general a non-square matrix. 
	In the following, we describe how we specify the measurements and tomography operations to generate the tomography matrix $\vb{T}$ for qubit and cavity tomography

	\subsubsection{Communication qubit state tomography}
	For tomography on the two communication qubits, which are physically transmon qubits, it is convenient to decompose the state in the Pauli basis: $\RhoOp = \sum_a p_a \cdot \hat{\sigma}_a$ where $\hat{\sigma}_a \in \lbrace \hat{\sigma}_I, \hat{\sigma}_x, \hat{\sigma}_y, \hat{\sigma}_z \rbrace^{\otimes N_q}$ are the generalized Pauli operators for an $N_q$-qubit quantum state.
	We then choose the overcomplete set of single-qubit rotations 
	$\lbrace \hat{I}, \hat{R}_x(\pi), \hat{R}_x(\pm\pi/2), \hat{R}_y(\pm\pi/2) \rbrace^{\otimes N_q}$ as tomography operations.
	Experimentally, we have the capability to measure the $\hat{\sigma}_z$ operator of each transmon qubit independently, thus extracting two bits of information for each experiment. 
	Ideally, this generates the set of computational-state projection operators: $\lbrace \hat{\Pi}_{00}, \hat{\Pi}_{01}, \hat{\Pi}_{10}, \hat{\Pi}_{11} \rbrace$, where $\hat{\Pi}_{jk} = \op{jk}$.
	In practice, we calibrate the measurement operators by preparing each of the four computational states and performing our two-bit measurement. The experimental POVM elements $\lbrace \hat{P}_{jk} \rbrace$ are then given as $\hat{P}_{jk} = \textrm{diag} \left[ Pr( 00 | \ket{jk}), Pr( 01 | \ket{jk}), Pr( 10 | \ket{jk}), Pr( 11 | \ket{jk}) \right]$, that is: the frequency of a particular measurement outcome given a particular state preparation, $\ket{jk}$.
	This analysis assumes that the measurement operator is only sensitive to $\hat{\sigma}_z$ component of the qubit state, and from previous work performing quantum detector tomography \cite{Blumoff2016}, we find this a reasonable assumption.
	
	\subsubsection{Data qubit state tomography} \label{sec:logical_qst}
	Reliable state tomography is predicated on ensuring small state preparation and measurement errors. However, when considering tomography on the data qubits (and in contrast to tomography on the communication qubits), it is no longer the case that we have a set of trusted operations to effect necessary operations and measurements on these multi-level systems. Therefore, we perform an indirect characterization \cite{Heeres2017} of logical qubit operations $\hat{U}_\text{op}$ where we perform tomography on the communication qubits for the composite operation $\hat{U}_\text{dec} \hat{U}_\text{op} \hat{U}_\text{enc}$. The protocol begins and ends in the communication qubit subspace and allows the use of trusted operation and measurements on the communication qubits.
	
	\subsubsection{Cavity Wigner tomography} \label{sec:cav_wigner_tomo}
	In the case for tomography of the cavity state, we measure the Wigner function by preparing the unknown cavity state $\RhoOp$, performing a set of displacements $\hat{D}_k \equiv \hat{D}(\beta{k}$, and measuring the photon-number parity $\Pi = \exp\left[ i \pi \hat{a}^\dagger \hat{a} \right]$, which leads to 
	\begin{equation} \label{eq:tomo_parity}
	W_k = \frac{2}{\pi} \Tr \left[ \RhoOp \hat{D}_k \hat{\Pi} \hat{D}_k^\dagger \right] 
	\end{equation}
	Next, we describe an efficient algorithm to extract the Wigner function for a given $\RhoOp$.
	We decompose $\RhoOp$ in the Fock-basis: $\RhoOp = \sum_{m,n}^{N_c} \rho_{m,n} \ketbra{m}{n}$, truncating at a maximum photon number $N_c$, and applying to \autoref{eq:tomo_parity}:
	\begin{equation} \label{eq:tomo_parity2}
	\begin{aligned}
	W_k &= \frac{2}{\pi} \sum_{m,n} \rho_{m,n} \Tr \left[ \op{m}{n} \hat{D}_k \hat{\Pi} \hat{D}_k^\dagger \right] \\
	&= \frac{2}{\pi} \sum_{m,n} \rho_{m,n} \bra{n}  \hat{D}_k^\dagger \Pi \hat{D}_k \ket{m}  \\
	&= \frac{2}{\pi} \sum_{m,n} \rho_{m,n}  W_{m,n}(\beta_k)
	\end{aligned}
	\end{equation}
	The matrix elements $W_{m,n}(\beta)$ can be efficiently calculated using the relation \cite{Leondart}
	\begin{equation}
	\begin{aligned}
	W_{m,n}(\beta) &\equiv \bra{n} \hat{D}(\beta) \Pi \hat{D}(\beta)^\dagger \ket{m} \\
	&= (-1)^m e^{-|\beta|^2} (2 \beta)^{m-n} \sqrt{\frac{n\!}{m\!}} L_n^{(n-m)}(|\beta|),
	\end{aligned}
	\end{equation}
	where $L_n^{(m-n)}$ is a generalized Laguerre polynomial.
	Thus, the tomography matrix elements has elements $\vb{T}_{[k,mn]} = W_{m,n}(\beta_k)$.
	
	\subsection{Reconstruction techniques}
	\subsubsection{State reconstruction}
	Though it is possible to directly calculate the density operator from inverting $\vb{T}$, noise and other experimental imperfections can result in unphysical reconstructed $\RhoOp$, possibly violating one or more of the requirements of non-negative eigenvalues, Hermeticity, or unit trace.
	We address this issue by utilizing a Maximum Likelihood Estimation (MLE) fit to the data.
	Given Gaussian statistics, the probability to observe the experimental measurement outcomes $\lbrace f_m \rbrace$ for a given $\rho$ is given by
	\begin{equation} \label{eq:tomo_mle_gaussian}
	\mathcal{L}\left( \lbrace f_m \rbrace | \rho \right) = 
	\frac{1}{\mathcal{N}} \prod_{m} \exp \left[ -\frac{\left[ \pi_m(\rho) - f_m \right]^2 }{2 \sigma_m^2}  \right],
	\end{equation}
	where $\pi_m$ is the expected value for a given experimental setting (given generally in \autoref{eq:tomo_eqn_general1}).
	We are interested in maximizing this probability by performing a search over all physical $\rho$. 
	To simplify this problem, we consider the logarithm of \autoref{eq:tomo_mle_gaussian} and assume the data are sampled from independent and identical distributions, $\sigma_r^m \rightarrow \sigma$. Given these we have the following residual sum-of-squares equation
	\begin{equation} \label{eq:tomo_mle_leastsq}
	\ln\mathcal{L}\left( \lbrace f_m \rbrace | \rho \right) = 
	\sum_{m} \left[ \pi_m(\rho) - f_m \right]^2,
	\end{equation}
	where we have dropped the negative sign to emphasize that this function is strictly non-negative and convex.
	Our reconstruction minimizes the log-likelihood function, \autoref{eq:tomo_mle_leastsq}.
	In order to specify a physical $\rho$, we specify this problem as a constrained semi-definite program, using CVXPY \cite{Grant2008,Grant2014} to solve the convex optimization problem
	\begin{equation}
	\begin{aligned}
	\rho_{MLE} = &\arg\!\min_\rho \quad \ln\mathcal{L}\left( \lbrace f_m \rbrace | \rho \right) \\
	& \text{subject to} \quad \rho \geq 0, \rho^\dagger = \rho, \Tr \rho = 1
	\end{aligned}
	\end{equation}
	
	%
	%
	
	\subsubsection{Process reconstruction}
	In our work, we perform logical process tomography on the teleported \CNOT gate. Our approach requires performing QST on an complete set of two-qubit initial states; here, we choose an overcomplete set of 36 input states $\lbrace \ket{\pm Z_L}, \ket{\pm X_L}, \ket{\pm Y_L} \rbrace^{\otimes2}$. Experimentally, we apply the appropriate rotation on each communication qubit and use optimal control pulses to encode the state onto the data qubit. Then, we perform the teleported \CNOT gate and subsequently apply another optimal control pulse to decode each state in the data qubits onto the communication qubit. With the quantum state contained in the communication qubit, we then perform QST to extract the state. With this set of ideal input states and experimentally reconstructed output states, we perform an inversion to extract the process that maps input states to output states. We represent the reconstructed process using the Pauli transfer matrix, $\mathcal{R}_\CNOT$ which relates input, $\vec{P}_{in}$, and output, $\vec{P}_{out}$, states in the Pauli basis, $\vec{P}_{out} = \mathcal{R}_{\textsc{CNOT}} \vec{P}_{in}$ \cite{Chow2012}. In \autoref{fig:extended_qpt_binomial} and \autoref{fig:extended_qpt_fock}, we present reconstructed process matrices for the Binomial and Fock encoding, respectively. For each, the statistical error as extracted from a bootstrap analysis is less than $1\%$; in the Main Text, error bars represent an estimate of the run-to-run variation, which is on the order of ${~}2\%$. 
	
	\subsection{Figures of merit}
	
	In this work we use the following two measures for state and process fidelity:
	\begin{enumerate}
		\item Fidelity between two states, $\rho$ and $\sigma$ \cite{Gilchrist2005}
		\begin{equation}
		\mathcal{F}_\text{state}\left( \rho, \sigma \right) = \Tr \left(\sqrt{\rho^{1/2} \sigma \rho^{1/2}}\right)
		\end{equation}
		\item Fidelity between two processes, $\mathcal{R}_1, \mathcal{R}_2$ \cite{Chow2012}
		\begin{equation}
		\mathcal{F}_\text{process} \left( \mathcal{R}_1, \mathcal{R}_2 \right) = \frac{\Tr\left[\mathcal{R}_1^T \mathcal{R}_2\right]/d + 1}{d + 1},
		\end{equation}
		with $d = 2 n$ and $n$ is the number of qubits.
	\end{enumerate}
	We also use the standard formula to calculate concurrence as given in \cite{Wootters1998}.
	
	We also note that the calculated process fidelity above is similar to the average gate fidelity, generally defined for two processes $\mathcal{E}_1$ and $\mathcal{E}_2$ as $\mathcal{F}_\text{avg} \equiv \int \dd \psi \mathcal{F}_\text{state}\left( \mathcal{E}_1(\rho), \mathcal{E}_2(\rho) \right)$.

	\subsection{Effect of leakage}
	
	As introduced in \autoref{sec:local_ops}, errors during the teleported \CNOT operation generally manifest themselves as codespace leakage errors (e.g. data qubit states outside of the subspace $\lbrace \ket{2}, \left(\ket{0} + \ket{4} \right) / \sqrt{2} \rbrace$).
	Therefore, these are important errors to include.
	In our experiments, when the data qubit states are decoded back onto the communication qubit, these leakage states are also mapped onto the communication two-dimensional subspace. Importantly, we do not attempt to postselect to remove these leakage cases.
	However, the behavior of our optimal control operations are specified only within the encoded subspace; the behavior of the operation outside the codespace is unconstrained when they are numerically optimized.
	Thus, it is not immediately clear how data qubit leakage errors are manifest within the communication qubit subspace.
	We use time-domain simulations to gain insight into the effect of leakage errors on tomography and fidelity calculations.
	We act the decode operation on a set of data qubit input states that are outside of the logical codespace (e.g. $\ket{1}$ or $\left( \ket{0} - \ket{4}\right) / \sqrt{2}$) and analyze the resulting communication qubit state.
	We typically the final communication qubit state to be mixed with relative populations between $\op{g}$ and $\op{e}$ dependent on the particular input state. If these populations were equal, then leakage errors on the data qubit would be mapped to depolarizing-type errors on the transmon. For a given state outside the codespace, one should not expect this behavior; however, the ensemble behavior of errors in a particular protocol can approximate a depolarizing channel. 
	To gain insight into the amount of bias that data qubit state leakage will introduce into our experimental tomography, we have performed time-domain simulations of the experimental tomography protocol, comparing extracted fidelities to the expected quantity.
	In general, we find that the fidelities extracted from a decode-style tomography experiment are only slightly different than the expected process fidelity, typically higher by around $0.5\%$. This bias in the tomography is a relatively small effect when we also include the decoherence induced by the decode process itself, which introduces around $3\%$ infidelity. 
	As such, leakage errors have minimal effect on our tomography, and we do not attempt to account for this bias in our fidelity calculations. 

\end{singlespace}

\newpage
%

\end{document}